\documentclass[11pt]{article}
\usepackage{fancyhdr}
\usepackage{isomath}
\usepackage{amsmath}
\usepackage{amsbsy}
\usepackage{amssymb}
\usepackage{amscd}
\usepackage{amsfonts}
\usepackage{array}
\usepackage{graphicx}
\usepackage{verbatim}
\usepackage{euscript}
\usepackage{alltt}
\usepackage{stmaryrd}
\usepackage{relsize}
\usepackage{enumitem}
\usepackage[lofdepth,lotdepth]{subfig}
\usepackage[numbers]{natbib}
\usepackage{caption}
\usepackage{float}
\usepackage{multirow}
\usepackage{epsfig}
\usepackage{color}
\usepackage{xcolor}
\usepackage{soul}
\usepackage{comment}

\DeclareGraphicsExtensions{.eps,.pdf,.png}
\usepackage{amsmath}
\usepackage{amsbsy}
\usepackage{amssymb}
\usepackage{amscd}
\usepackage{amsfonts}

\newcommand{\bs}{\boldsymbol}

\newcommand{\bfa}{{\mathbold a}}
\newcommand{\bfb}{{\mathbold b}}
\newcommand{\bfc}{{\mathbold c}}
\newcommand{\bfd}{{\mathbold d}}
\newcommand{\bfe}{{\mathbold e}}
\newcommand{\bff}{{\mathbold f}}

\newcommand{\bfm}{{\mathbold m}}
\newcommand{\bfn}{{\mathbold n}}

\newcommand{\bfv}{{\mathbold v}}

\newcommand{\bfx}{{\mathbold x}}

\newcommand{\bfA}{{\mathbold A}}
\newcommand{\bfB}{{\mathbold B}}

\newcommand{\bfE}{{\mathbold E}}
\newcommand{\bfF}{{\mathbold F}}

\newcommand{\bfI}{{\mathbold I}}

\newcommand{\bfL}{{\mathbold L}}

\newcommand{\bfR}{{\mathbold R}}

\newcommand{\bfT}{{\mathbold T}}

\newcommand{\bfV}{{\mathbold V}}
\newcommand{\bfW}{{\mathbold W}}
\newcommand{\bfX}{{\mathbold X}}

\newcommand{\beq}{\begin{equation}}
\newcommand{\eeq}{\end{equation}}
\newcommand{\beqs}{\begin{eqnarray}}
\newcommand{\eeqs}{\end{eqnarray}}
\newcommand{\beql}{\begin{equation} \label}

\newcommand{\bfchi}{\mathbold{\chi}}

\newcommand{\bfalpha}{\mathbold{\alpha}}

\newcommand{\bfzero}{\mathbf{0}}

\newcommand{\grad}{\mathop{\rm grad}\nolimits}
\newcommand{\divergence}{\mathop{\rm div}\nolimits}
\newcommand{\curl}{\mathop{\rm curl}\nolimits}

\usepackage[margin=1in]{geometry}
\usepackage[finalnew]{trackchanges}

\date{}
\begin{document}
\title{Interface-dominated plasticity and kink bands in metallic nanolaminates}
\author{Abhishek Arora\thanks{Department of Civil \& Environmental Engineering, Carnegie Mellon University, Pittsburgh, PA 15213} $\qquad$  Rajat Arora\thanks{Advanced Micro Devices, Austin, TX 78735} $\qquad$  Amit Acharya\thanks{Department of Civil \& Environmental Engineering, and Center for Nonlinear Analysis, Carnegie Mellon University, Pittsburgh, PA 15213, email: acharyaamit@cmu.edu.}}
\maketitle

\begin{abstract}
 \noindent The theoretical and computational framework of finite deformation mesoscale field dislocation mechanics (MFDM) is used to understand the salient aspects of kink-band formation in Cu-Nb nano-metallic laminates (NMLs). A conceptually minimal, plane-strain idealization of the three-dimensional geometry, including crystalline orientation, of additively manufactured NML is used to model NMLs. Importantly, the natural jump/interface condition of MFDM imposing continuity of (certain components) of plastic strain rates across interfaces allows theory-driven `communication' of plastic flow across the laminate boundaries in our finite element implementation. Kink bands under layer parallel compression of NMLs in accord with experimental observations arise in our numerical simulations. The possible mechanisms for the formation and orientation of kink bands are discussed,  within the scope of our idealized framework. We also report results corresponding to various parametric studies that provide preliminary insights and clear questions for future work on understanding the intricate underlying mechanisms for the formation of kink bands. 
\end{abstract}

\section{Introduction} \label{introduction}
\begin{figure}[htbp]
    \centering
    \subfloat[][]{
    \includegraphics[scale=0.85]{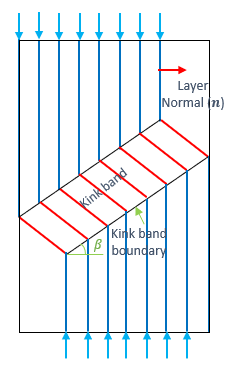}} \; \; \; \;
    \subfloat[][]{
    \includegraphics[scale=1.4]{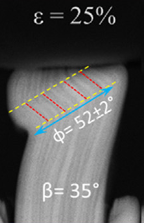}} 
    \caption{(a) Schematic of an idealized kink band in layer parallel compression of NML, (b) geometry of the kink band at $|e_N| = 25\%$ engineering strain in the experimental work of compression of Cu/Nb NMLs. (Figure in (b) reprinted from \cite{zhang2022kink} with permission from \textit{Elsevier.})}
    \label{fig:schematic_experimental_kink_band}
\end{figure}

A fundamental challenge in the development of efficient engineering materials is to design a material that has both high strength and high toughness, as most materials have exclusively either of these properties. Kinking in nano-metallic laminates is a deformation mechanism that may circumvent this challenge, as it leads to enhanced plasticity (ductility) at an almost constant level of stress after yielding \cite{beyerlein2022mechanical}. Kink bands, in the present context of a material with an intrinsic layered structure, refer to a thin band of localized deformation that is transverse - and most commonly not normal - to the intrinsic layering, as shown in Fig.~\ref{fig:schematic_experimental_kink_band} (a). Shear band formation has been the predominantly observed form of localization in NMLs as reported in \cite{mara2008deformability, beyerlein2013interface}. In these studies, NMLs are compressed perpendicular to the layer direction in micropillar configurations, \add{meaning the compression loading direction is parallel to the layer normal direction}. While this is a somewhat physically intuitive mode of deformation localization through rotation of the layers towards the loading axis and slipping along softer layers, kink band formation in NMLs for sufficiently small layer thickness under layer parallel compression has also been demonstrated \cite{nizolek2015enhanced}. \add{In layer parallel compression of NML, the compression loading direction is perpendicular to the layer normal direction as shown in Fig.~\mbox{\ref{fig:schematic_experimental_kink_band}} (a).} The formation of such kink bands and its dependence on the size of layers is quite non-intuitive, and it is the study of this phenomenon that forms the subject of this paper; the modeling of the development of shear bands in NMLs in micropillar configurations under layer perpendicular compression is a subject worthy of study on its own merits, a challenge that has not been successfully solved as yet \cite{zecevic2023non}.

In general, the formation of shear or kink bands, due to strain localization, is detrimental as it produces sites of narrow planes under intense shear or lattice rotations, which acts as a source for crack initiation and leads to eventual material failure. For instance, it was reported in \cite{jia2003effects} that for grain sizes smaller than 300 nm in iron (Fe), shear bands are formed leading to strain-softening and failure at low values of plastic strain. Similarly, graphite-fiber-reinforced epoxy composites fail under compression loading due to the formation of kink bands \cite{narayanan1999mechanisms}. However, for layer parallel compression of NMLs with sufficiently small laminate widths, a large amount of plastic deformation is observed before failure without much reduction in the strength of the material in the experimental work of Thomas et al.~\cite{nizolek2015enhanced} in Cu-Nb nanolaminates. It was reported there that the samples with layer thickness greater than 250 nm deform homogeneously, while the deformation behavior of samples with thicknesses 65 or 30 nm was inhomogeneous, and enhanced plastic deformation was observed after yield at a more-or-less constant load \cite{nizolek2015enhanced}. Recently, Zhang et al.~\cite{zhang2022kink} investigated the kink band formation process in Cu/Nb NMLs via \textit{in situ} micropillar compression along the layer-parallel direction in a scanning electron microscope (SEM). It was reported that after the onset of kinking, dislocations active on the slip systems nearly parallel to the interface in NMLs are responsible for the band evolution. Cahn \cite{cahn1950slip} showed the formation of kink bands in aluminum single crystals, and it was suggested that these bands are composed of dipolar arrays of edge dislocations aligned perpendicular to the primary slip direction.

With respect to modeling, Zecevic et al.~\cite{zecevic2023non} used Gurtin's \cite{gurtin2002gradient} strain gradient crystal plasticity model augmented with various choices of the defect energy function within a large strain elasto-viscoplastic Fast Fourier transform-based computational technique to model the compression of Cu/Nb NMLs.  Arora and Acharya \cite{arora2020dislocation} modeled a 2-D single crystal under simple shearing using the mesoscale field dislocation mechanics (MFDM) framework in a finite deformation setting. In a special single slip study, kink bands were observed in the form of dipolar dislocation walls perpendicular to the single slip plane \cite{arora2020dislocation}. \add{Kink band formation in a polycrystalline Mg alloy under compression was computationally studied using a  strain-gradient crystal plasticity framework \mbox{\cite{ueta2015dislocation}}.} Asaro and Rice \cite{asaro1977strain} described the onset of a shear and kink band from a bifurcation analysis of the classical, finite deformation, single slip constitutive behavior. Forest \cite{forest1998modeling} used small deformation generalized crystal plasticity (Cosserat theory of crystal plasticity) to separate bifurcation modes (slip and kink bands), i.e. kink bands occur later than slip bands, and this is unlike the prediction of classical plasticity.

In this work, we model the compression of composite nanolaminates within a  plane strain idealization, using the theoretical and computational framework of MFDM. The interface conditions in MFDM impose continuity of certain components of the total plastic strain rate, which models continuous plastic flow across the laminate interfaces. We consider a `minimal' model setup in considering 3 co-planar slip systems for each material in the bi-metallic laminate structure and show that kink bands are formed in qualitative accord with the experimental observations of \cite{nizolek2015enhanced, zhang2022kink}. We also report on other studies within this idealized setup and discuss whether kink bands form or not for different cases.

The outline of the paper is as follows: after presenting some notational details used in this paper in the following paragraph, Section \ref{theory} provides a brief review of the theory of MFDM. For more details on the theoretical and computational framework of MFDM, the interested reader is referred to \cite{arora2020dislocation,arora2020finite,arora2020unification,arora2019computational}. Section \ref{review_of_results_mfdm} discusses the various successes of the MFDM framework with correspondence with experimental studies in many instances. Section \ref{results_and_discussion} shows the numerical results of kink-band formation in composite nanolaminates within our idealization using the computational framework of MFDM. We also discuss various parametric studies of our model characterizing this complicated problem of kink band formations in NMLs. Section \ref{conclusion} presents some concluding remarks and the scope of future work for modeling compression of NMLs using our framework.

Vectors and tensors are represented by boldface lower and upper-case letters, respectively. The action of a second-order tensor $\bfA$ on a vector $\bfb$ is denoted by $\bfA \bfb$. The inner product of two vectors is denoted by $\bfa \cdot \bfb$, while the inner product of two second-order tensors is denoted by $\bfA:\bfB$. A rectangular Cartesian coordinate system is invoked for the ambient space and all vector and tensor components are expressed with respect to the basis of this coordinate system. The partial derivative of the quantity $(\cdot)$ w.r.t.\ the $x_i$ coordinate of this coordinate system is denoted by $(\cdot)_{,i}$, and $\bfe_i$ denotes the unit vector in the $x_i$ direction. The time derivative of a quantity $(\cdot)$ is denoted by $\dot{(\cdot)}$. Einstein's summation convention is always implied unless mentioned otherwise. The symbols $\grad$, $\divergence$, and $\curl$ denote the gradient, divergence, and curl on the current configuration, respectively. For a second order tensor $\bfA$, vectors $\bfv$, $\bfa$, and $\bfc$, and a spatially constant vector field $\bfb$, the operations of $\divergence$, $\curl$, and the cross-product of a tensor $(\times)$ with a vector are defined as follows: 
\begin{subequations}
\begin{align*}
    \left( \divergence \bfA \right) \cdot \bfb = \divergence \left( \bfA^T \bfb \right),& \qquad \forall \:\: \bfb \\
    \bfb \cdot \left( \curl \bfA \right) \bfc = \left[ \curl\left(\bfA^T \bfb \right)\right] \cdot \bfc,& \qquad \forall \: \: \bfb, \bfc \\
    \bfc \cdot \left( \bfA \times \bfv \right) \bfa = \left[ \left(\bfA^T \bfc \right) \times \bfv \right] \cdot \bfa,& \qquad \forall \: \: \bfa, \bfc.
\end{align*}
\end{subequations}
In rectangular Cartesian coordinates, these are denoted by
\begin{subequations}
\begin{align*}
    \left( \divergence \bfA \right)_i =  A_{ij,j}, \quad
    \left( \curl \bfA \right)_{ri} = \varepsilon_{ijk} A_{rk,j}, \quad
    \left( \bfA \times \bfv \right)_{ri} = \varepsilon_{ijk} A_{rj} v_{k},
\end{align*}
\end{subequations}
where $\varepsilon_{ijk}$ are the components of the third order alternating tensor $\bfX$. 

When we refer to different slip systems with respect to their slip vector orientation, we mean that those orientations were in the initial unstretched lattice, and do not represent the orientations of the corresponding slip vectors in the deformed lattice, unless mentioned otherwise.
\section{Review of \add{MFDM} theory}\label{theory}
The governing equations for MFDM \citep{arora2020finite} are as follows:
\begin{subequations}\label{mfdm_short} 
\begin{gather}
    \curl \bs{W} = - \bs{\alpha}, \label{incompatibility_W}\\ 
    \dot{\bfW} + \bfW \bfL = \bs{\alpha} \times \bfV,  \label{conservation_law_W} \\
    \divergence \left[ \bfT (\bfW) \right] = \bigg\{ \begin{array}{l}
    \bs{0}, \quad \text{quasistatic} \\
    \rho \dot{\bfv}, \quad \text{dynamic}.
    \end{array}  \label{equilibrium}
\end{gather} 
\end{subequations}  
In Eq.~\eqref{mfdm_short}, $\bs{W}$ is the inverse of the elastic distortion tensor ($\bs{F}^e$), $\bfalpha$ denotes the dislocation density tensor \add{(Nye's tensor)}, $\bfv$ denotes the material velocity, $\bfL = \grad \bfv$ is the velocity gradient, $\bfT$ is the (symmetric) Cauchy stress tensor\add{, and $\bfV$ denotes the dislocation velocity field with respect to the material}. Eq.~\eqref{incompatibility_W} is the incompatibility equation for the inverse elastic distortion tensor, while Eq.~\eqref{conservation_law_W} is derived from the statement of conservation of \add{the} Burgers vector content of an arbitrary area patch in the current configuration \cite{acharya2011microcanonical}. Eq.~\eqref{equilibrium} is the linear momentum balance for both quasistatic and dynamic cases.

Eq.~\eqref{conservation_law_W} is a system of coupled Hamilton-Jacobi equations for which, to our knowledge, no rigorous mathematical theory or numerical schemes exist in the literature. To computationally approximate the governing equations, a \change{Stokes-Helmoholtz}{Stokes-Helmholtz} decomposition of the field $\bfW$ is introduced, expressing it as a sum of a gradient $(\grad \bff)$ and a divergence-free part ($\bs{\chi}$) \citep{arora2020dislocation}. This decomposition replaces the tensor field $\bfW$ with a tensor and a vector field in our computations, but it is relatively easier to develop numerical schemes for this larger set of governing equations \cite{arora2020finite}. 

The extended \add{set of} governing equations for MFDM \cite{arora2020finite} are shown below:
\begin{subequations}\label{eqns_set:mfdm}
\begin{gather}
     \mathring{\bs{\alpha}} \equiv (\divergence(\bs{v})) \bs{\alpha} + \dot{\bs{\alpha}} - \bs{\alpha} \bfL^T = - \curl (\bs{\alpha} \times \bfV +  \bfL^p), \label{alpha_conservation} \\
    \begin{array}{c} 
      \bfW = \bs{\chi} + \grad \bff, \: \bs{F}^e = \bs{W}^{-1}, \\
      \curl \bs{\chi} = - \bs{\alpha},  \\
    \divergence \bs{\chi}  = \bs{0},         
    \end{array} \Bigg \} \label{incompatibility} \\ 
    \divergence \left( \grad \dot{\bs{f}} \right) = \divergence \left( \bs{\alpha} \times \bs{V} +  \bfL^p - \dot{\bs{\chi}} - \bs{\chi} \bs{L} \right), \label{plastic_eq} \\
    \divergence \left[ \bfT (\bfW) \right] = \bigg\{ \begin{array}{l}
    \bs{0}, \quad \text{quasistatic} \\
    \rho \dot{\bfv}, \quad \text{dynamic}.
    \end{array}  
\end{gather}
\end{subequations}
Here, \eqref{alpha_conservation} is the evolution equation for the dislocation density tensor, and \eqref{incompatibility} is the incompatibility equation in terms of $\bs{\chi}$ with a condition that $\divergence \bs{\chi} = \bs{0}$, while $\bff$ denotes the plastic position vector, and the evolution equation for $\bff$ is given by \eqref{plastic_eq}. $\bfL^p$ is the additional mesoscale field which represents the averaged rate of plastic straining due to all dislocations that cannot be represented by $\bfalpha \times \bfV$, where both fields in the product represent space-time running averages \add{of the corresponding microscopic fields}. \add{This is due to the fact that the average of products is not the same as the product of averages, in general.} \change{except for $\bfL^p$}{Hence,} all fields in the MFDM framework are running space-time averages of \add{the} corresponding fields of \change{FDM}{the field dislocation mechanics (FDM)} theory\add{; except for $\bfL^p$, as there is no corresponding field in FDM theory to $\bfL^p$.}

To close the above set of equations in \eqref{eqns_set:mfdm}, constitutive statements for $\bs{V}, \bs{L}^p, \bs{T}$ are used, consistent with the mechanical dissipation being non-negative \citep{arora2020dislocation}.

Initial conditions for $\bfalpha, \bff$ and boundary conditions for $\bfalpha, \bff, \bfchi$, and $\bfv$ are specified, for a  well-set evolution in MFDM.

\subsection{Boundary conditions}
\begin{itemize}
    \item The $\bs{\alpha}$ evolution equation (shown in Eq.~\eqref{alpha_conservation}) has a convective boundary condition of the form $(\bfalpha \times \bfV + \bfL^p) \times \bfn = \bs{\Phi}$, where $\bs{\Phi}$ is a second order tensor valued function of position and time on the boundary, characterizing the flux of dislocations at the surface satisfying the constraint $\bs{\Phi} \bfn = \bfzero$. Here, $\bfn$ isThe outward unit normal field on the boundary.
    
    \remove{There are two ways in which}The boundary condition is specified \add{in two ways}: (a) \textit{Plastically constrained}: $\bs{\Phi} (\bfx, t) = \bfzero$ \add{is specified} at a point $\bfx$ on the boundary for all times, which ensures that there is no outflow of dislocations at that point of the boundary, and only parallel motion along the boundary is allowed. (b) \textit{Plastically unconstrained}: A less restrictive boundary condition where $\hat{\bfL}^p \times \bfn$ is simply evaluated at the boundary (akin to an outflow condition), along with the specification of dislocation flux $\bfalpha (\bfV \cdot \bfn)$ on the inflow part of the boundary. Additionally, for all calculations presented in this paper $(\curl \bfalpha \times \bfn) = \bf0$ is imposed, a particular specification of a boundary condition that arises from simple mathematical modeling of the manifestation of dislocation core energy at the mesoscale.
    \item For the incompatibility equation, $\bs{\chi} \bs{n} = \bfzero$ is applied on the outer boundary of the domain, which along with the system \eqref{incompatibility} ensures that $\bfchi$ vanishes when $\bfalpha$ is zero in the entire domain. 
    \item The $\bs{f}$ evolution equation requires a Neumann boundary condition i.e., $(\grad \dot{\bff}) \bfn = (\bfalpha \times \bfV + \bfL^p - \dot{\bfchi} - \bfchi \bfL ) \bfn$ on the outer boundary of the domain.
    \item The material velocity boundary conditions are applied based on the loading type, which is discussed later in Section \ref{results_and_discussion}. 
\end{itemize}

\subsection{Initial conditions}
\begin{itemize}
    \item The initial condition $\bs{\alpha} (\bs{x}, 0) = \bs{0}$ is assumed for all simulations here. 
    \item In general, the initial condition for $\bs{f}$ is obtained by solving for $\bfchi$ from the incompatibility equation and solving for $\bs{f}$ from the equilibrium equation, for prescribed $\bs{\alpha}$ on the given initial  configuration. We refer to this scheme as the elastic theory of continuously distributed dislocations (ECDD). For the initial conditions on $\bfalpha$ considered above, this step is trivial, with $\bff = \bfX$, where $\bfX$ is the position field on the initial configuration.
    \item The model admits an arbitrary specification of $\dot{\bff}$ at a point to uniquely evolve $\bff$ using \eqref{plastic_eq} in time, and this rate is prescribed to vanish.
\end{itemize}

\noindent \textit{Interface conditions for plastic distortion rate}: The interface jump condition for the $\bfalpha$ solve is $ [[\bfalpha \times \bfV + \hat{\bfL}^p + l^2 \hat{\gamma} \curl \bfalpha ]] \times \bfn = \bf0$, and this gets imposed in the finite element formulation. \add{Here, $[[(\cdot)]]$ denotes the jump in $(\cdot)$ across the interface.} This \add{interface condition} allows for the discontinuity of certain components of plastic strain rate across the interface \cite{acharya2007jump}.

Traction continuity and natural interface conditions consistent with the PDE for $\bfchi, \dot{\bff}$ are also enforced by our finite element formulation.

\subsection{Constitutive relations}
Constitutive relations in MFDM are required for the stress $\bfT$, the plastic distortion rate $\bfL^p$, and the dislocation velocity $\bfV$. The details of the thermodynamically consistent constitutive formulations can be found in Sec.~3.1 of \cite{arora2020dislocation}. 

Table \ref{tab:constitutive_relation_T} presents the constitutive relation for Cauchy stress and mesoscopic core energy density for the material. Here, $\rho^*$ denotes the mass density of the pure, un-stretched elastic lattice, and $\mathbb{C}$ is the fourth-order elasticity tensor, while $\bfE^e$ is the elastic Lagrangian strain tensor. $\Upsilon (\bfalpha)$ is a heuristic representation of the averaged microscopic core energy at the mesoscale, and $\epsilon$ is a constant that has physical dimensions of $stress \times {length}^2$.
{\renewcommand{\arraystretch}{1.8}
\begin{table}[htbp]
    \centering
    \begin{tabular}{l c}
    \hline 
        Saint-Venant-Kirchhoff Material & 
            \(\displaystyle 
                \phi(\bfW) = \frac{1}{2 \rho^*} \bfE^e : \mathbb{C} : \bfE^e,  \quad \bfT = \bfF^e [\mathbb{C} : \bfE^e] {\bfF^e}^T 
              \) \\    \hline 
         Core energy density &
          \(\displaystyle
              \Upsilon (\bfalpha) = \frac{1}{2 \rho^*} \epsilon \bfalpha : \bfalpha
          \) \\ 
    \hline
    \end{tabular}
    \caption{Constitutive relations for Cauchy stress and core energy density.}
    \label{tab:constitutive_relation_T}
\end{table}}

Table \ref{tab:constitutive_relation_Lp} shows the constitutive relations for plastic distortion rate. Classical crystal plasticity is assumed here and the plastic distortion $\bfL^p$ is given by the sum of slipping distortions on prescribed slip systems \cite{arora2020dislocation}. In Table \ref{tab:constitutive_relation_Lp}, $(\cdot)_{sym}$ denotes the symmetric part of $(\cdot)$, $\hat{\gamma}_0^k$ is a reference strain rate, $\hat{\gamma}^k$ is the slipping rate due to statistically stored dislocations on the $k^{th}$ slip system, $n_{sl}$ is the total number of slip system, $sgn(\tau^k)$ denotes the sign of the scalar $\tau^k$, $g$ is the material strength, $m$ is the material rate sensitivity. $\bfm^k$ and $\bfn^k$ represents the slip direction and slip plane normal for the $k^{th}$ slip system, in the current configuration. The resolved shear stress on the $k^{th}$ slip system is given as $\tau^k = \bfm^k \cdot \bfT \bfn^k$. The slip direction and slip plane normal in the current configuration are related to the unstretched slip direction ($\bfm^k_0$) and the slip plane normal ($\bfn^k_0$) by $\bfm^k = \bfF^e \bfm^k_0$, and $\bfn^k = {\bfF^e}^{-T} \bfn^k_0$. The classical plastic distortion rate $\hat{\bfL}^p$ from crystal plasticity, is augmented with an additive term in $\curl \bfalpha$, motivated from the mechanical dissipation expression in \cite[Eq.~(8)]{arora2020dislocation}, with a mobility coefficient given by $l^2$ multiplied by the average slipping rate of the total slip systems \cite{arora2020dislocation}. Here, $l$ is a material length related to the gross modeling of mesoscale effects of dislocation core energy and is defined as $l^2 = \epsilon/g_0$, with $g_0$ being the  initial strength (initial yield stress in shear).

{\renewcommand{\arraystretch}{1.8}
\begin{table}[htbp]
    \centering
    \begin{tabular}{l c}
    \hline 
        \multirow{2}{6em}{Crystal plasticity}  & 
            \(\displaystyle
                \hat{\bfL}^p = \bfW \left( \sum_{k}^{n_{sl}} \hat{\gamma}^k \bfm^k \otimes \bfn^k \right)_{\textit{sym}}; \quad 
                \hat{\gamma}^k = \textit{sgn}(\tau^k) \hat{\gamma}_0^k \left( \frac{|\tau^k|}{g} \right)^{\frac{1}{m}}
            \) \\ 
            &  \(\displaystyle \quad \bfL^p = \hat{\bfL}^p + \frac{l^2}{n_{sl}} \sum_{k}^{n_{sl}} |\hat{\gamma}^k| \curl \bfalpha \) \label{lp_equation} \\
    \hline
    \end{tabular}
    \caption{Constitutive relations for plastic strain rate $\bfL^p$.}
    \label{tab:constitutive_relation_Lp}
\end{table}}
Table \ref{tab:constitutive_relation_V} shows the constitutive relations for dislocation velocity. \add{Here, $\bfd$ denotes the direction of dislocation velocity, and its magnitude is given by $\zeta$, $\bfT'$ denotes the deviatoric stress tensor,} $\bfX$ is the third-order alternating unit tensor, $\mu$ is the shear modulus, $g$ is the material strength, $\eta$ is a non-dimensional material constant in the empirical Taylor relationship for macroscopic strength vs dislocation density, and $b$ is the Burgers vector magnitude of a full dislocation in the crystalline material. \add{For motivation behind the constitutive statement for $\bfV$, the interested reader is referred to \mbox{\cite{acharya2006size,acharya2012elementary}}.}

{\renewcommand{\arraystretch}{1.8}
\begin{table}[htbp]
    \centering
    \begin{tabular}{c c c}
    \hline 
        \(\displaystyle \bfT^{'} = \bfT - \frac{tr(\bfT)}{3} \bfI\);
        &  \(\displaystyle
            \bfa = \frac{1}{3} \bfX (tr(\bfT)  \bfF^{e} \bfalpha );
            \)    
            &  \(\displaystyle \bfc = \bfX( \bfT^{'} \bfF^{e} \bfalpha ) \)  \\
    \( \displaystyle\bfd = \bfc - \left( \bfc - \frac{\bfa}{|\bfa|} \right) \frac{\bfa}{|\bfa|}; \) 
    & \(\displaystyle \bfV = \zeta \frac{\bfd}{|\bfd|} ; \)  
    & \(\displaystyle \zeta = \frac{\mu^2 \eta^2 b}{g^2 n_{sl}} \: \sum_{k}^{n_{sl}} |\hat{\gamma}^k|\) \\
    \hline
    \end{tabular}
    \caption{Constitutive relations for dislocation velocity $\bfV$.}
    \label{tab:constitutive_relation_V}
\end{table}}
Table \ref{tab:constitutive_relation_g} shows the evolution equation for material strength ($g$). Here, $g_s$ denotes the  saturation strength, $h$ is the total hardening rate, $\Theta_0$ is the Stage II hardening rate, $k_0$ (non-dimensional) characterizes the hardening rate due to geometrically necessary dislocations (GNDs) \cite{acharya2000grain}.
{\renewcommand{\arraystretch}{1.8}
\begin{table}[H]
    \centering
    \begin{tabular}{c c}
    \hline 
    \( \displaystyle \dot{g} = h(\bs{\alpha}, g) \left(|\bs{F}^e \bs{\alpha} \times \bs{V}| + \sum_{k}^{n_{sl}} |\hat{\gamma}^k| \right); \) 
    & \( \displaystyle h(\bs{\alpha}, g) =  \frac{\mu^2 \eta^2 b}{2 (g-g_0)} k_{0} |\bs{\alpha}| + \Theta_{0} \left(\frac{g_s-g}{g_s - g_0} \right) \)  \\
    \hline
    \end{tabular}
    \caption{Constitutive relations for material strength $g$.}
    \label{tab:constitutive_relation_g}
\end{table}}
All parameters in our model, except $k_0$ and $l$, are part of the constitutive structure of well-accepted models of classical crystal plasticity theory. The parameter $k_0$ was introduced in \cite{acharya2000grain}. The length scale $l$ simply controls the refinement of the GND microstructure and does not play a physically significant role in our results.

\section{Review of results from MFDM} \label{review_of_results_mfdm}
The MFDM framework is based on the space-time averaging of microscopic dislocation mechanics, and the obtained governing PDEs shown in \eqref{eqns_set:mfdm} are closed by the phenomenological specification of $\bfL^p$, $\bfV$, $\bfT$ and $g$. This approach has been quite successful in addressing some challenging problems in modern plasticity theory related to the computation of patterning \cite{arora2020dislocation}, dislocation internal stress \cite{arora2020finite}, size effects in micropillar confined thin metal films \cite{arora2022mechanics}, polygonization \cite{arora2020unification}, and slip transmission at grain boundaries \cite{puri2011controlling, puri2011mechanical} among others \cite{acharya2006size, acharya2007jump, puri2010modeling, mach2010continuity, fressengeas2010dislocation, acharya2011microcanonical, das2016microstructure, fressengeas2009dislocation, taupin2007effects, taupin2010particle, richeton2011continuity, taupin2008directionality, djaka2015numerical, varadhan2009lattice, djaka2020fft, berbenni2020fast, genee2020particle}, including long-standing and recent fundamental challenges in the prediction of large-deformation, dislocation mediated elastic and elastic-plastic response \cite{arora2020dislocation,arora2020finite,arora2020unification, arora2022mechanics}. The benefits of using such a framework are:
\begin{itemize}
    \item It is difficult to deduce a physical connection between the plastic strain/distortion in classical plasticity theory to the mechanics of dislocation, beyond modeling in 1-D. The MFDM framework brings out an explicit connection between the plastic strain rate and the motion and geometry of an evolving microscopic array of dislocations. Obviously, this has many benefits, even for a phenomenological specification of the macroscopic plastic strain rate. Moreover, the MFDM framework has allowed for a first unification between phenomenological $\textit{J}_2$ and crystal plasticity theories and quantitative dislocation mechanics.
    
    \item With a single extra material fitting parameter beyond a classical plasticity model (and two in the finite deformation setting), the MFDM framework has \change{enables}{enabled} a significant variety of phenomena to be modeled, in qualitative and quantitative accord with experimental results. 
\end{itemize}

\section{Results and Discussion} \label{results_and_discussion}
The initial configuration of NMLs, as shown in Fig.~\ref{fig:stress-strain-curve-comparison} (a),  is such that there are 12 alternating layers of Cu and Nb. Each layer has a thickness of $l = 80 \: nm$ with the width of the composite being $W = 960 \: nm$. The aspect ratio of the NML composite is 1:2, with an undeformed height of the sample being $L = 1920 \: nm$. The nominal compression loading rate is $|\hat{e}| = 0.001\, s^{-1}$, and at any time $t$, the nominal compressive strain is $|e| = |\hat{e}| t$. The corresponding true compressive strain is denoted as $|e_{T}|$. The average true compressive stress is denoted by $\sigma$, while the average nominal compressive stress is denoted by $\sigma_N$. Plastically unconstrained boundary conditions are applied to all four (external) boundaries of the domain, for all results shown in this work. Quasi-static force equilibrium is solved for all simulations reported here.

The boundary conditions for the material velocity for the simulations of compression of NMLs are as follows: 
\begin{itemize}
    \item $v_2 = 0$ at the bottom boundary of the domain.
    \item $v_2 = -|\hat{e}| L$ at the top boundary of the domain, where $L$ is the length of the laminate in the (undeformed) initial configuration at $t=0$.
    \item The applied traction in the horizontal direction is zero on both the top and the bottom boundary of the domain.
    \item $v_1 = 0$ at a single node of the bottom boundary. This along with the above conditions suffices to constrain rigid motion.
\end{itemize} 

As discussed earlier, we assume 3 slip systems for both Cu and Nb. One of the reference slip directions is aligned along the layer parallel direction, while the other two are defined to initially accommodate symmetric double slip with respect to the loading axis. With reference to Fig.~\ref{fig:stress-strain-curve-comparison} (a), $\theta_1$ and $\theta_2$ are both assumed to be $30^{\circ}$ for Cu, while for Nb, both are $45^{\circ}$, unless mentioned otherwise. 

Table \ref{table:material_properties} shows the material parameters used for the numerical simulations of NMLs under compression in this study. The $g_0$ values for Cu and Nb are taken from \cite{zecevic2023non}, and the saturation strength is assumed to be $g_s = 9 g_0$. We use these material properties for all of our simulations unless otherwise stated, although these values of initial and saturation yield strength for single crystal materials seem somewhat out of the norm. Accordingly, we also perform simulations with more commonly used values without a qualitative change in results. 

\begin{table}[htbp]
\centering
\begin{tabular}{ |c| c c c c c c c c c c c|} 
 \hline
 \multirow{2}{4.5em}{Parameter} & $\hat{\gamma}_0$ & $m$ & $\eta$ & $b$ & $g_0$  & $g_s$  &  $\Theta_0$  & $k_0$ & $l$ & $E$ & $\nu$ \\ 
   & ($s^{-1}$) &   &  & (\AA) & (MPa) & (MPa) &  (MPa) & & ($\mu m$) & (GPa) &  \\ 
 \hline
  Cu & 0.001  & 0.03 & $\frac{1}{3}$ & 2.556 & 210   & 1890   & 273  & 20  & $\sqrt{3} \times 0.1$  & 144.58 & 0.324 \\
  \hline
  Nb & 0.001  & 0.03 & $\frac{1}{3}$ & 2.86 & 262.5   & 2362.5  & 198  & 20  & $\sqrt{3} \times 0.1$  & 110.25 & 0.392 \\
 \hline
\end{tabular}
\caption{Material parameters for Cu and Nb.}
\label{table:material_properties}
\end{table}

\subsection{Kink band formation}

\begin{figure}[htbp]
    \centering
     \subfloat[][]{
    \includegraphics[scale=0.55]{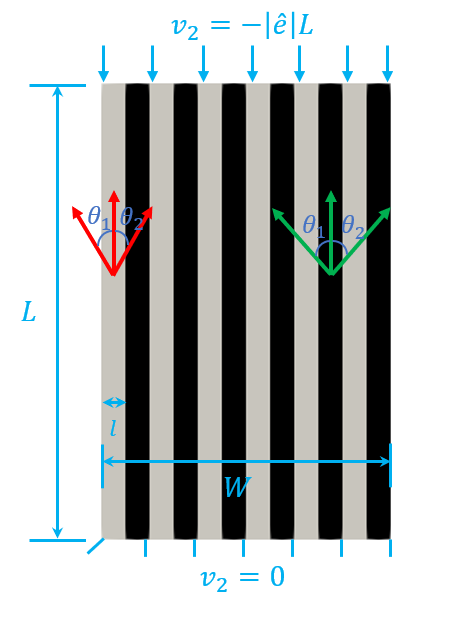}}
    \subfloat[][]{
    \includegraphics[scale=0.4]{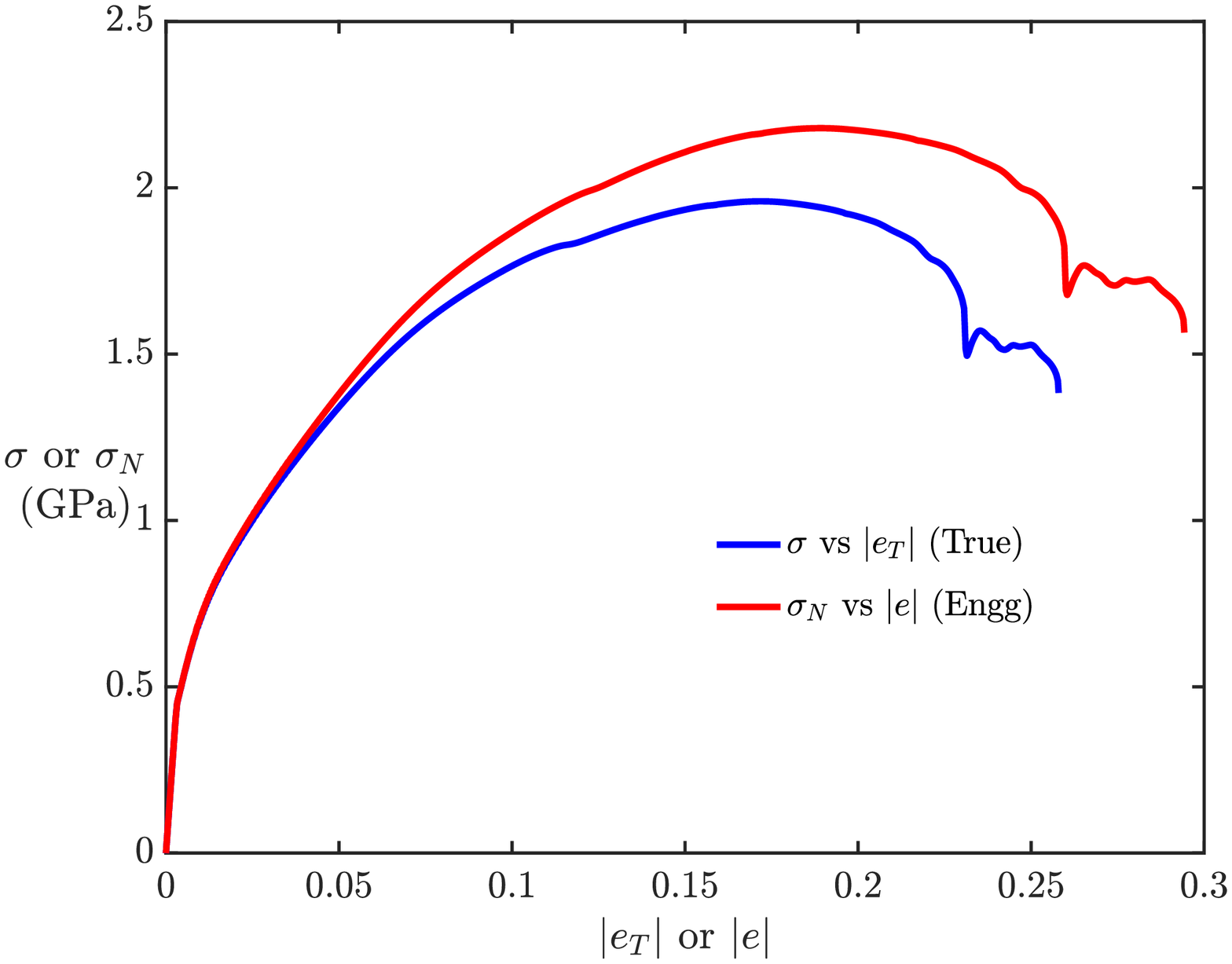}} \quad
    \subfloat[][]{
    \includegraphics[scale=0.6]{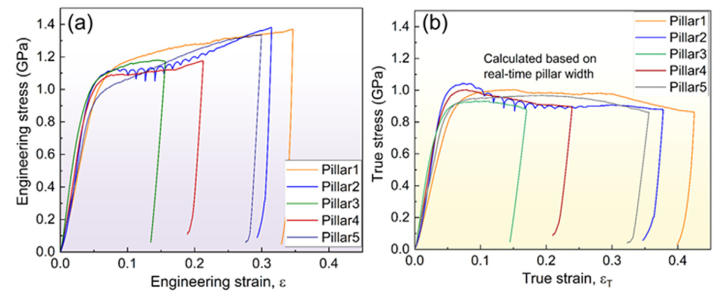}}
    \caption{(a) Schematic of NML under compression with Cu layers shown in grey color and Nb layers in black color, (b) stress-strain curves (both true and engineering) from our simulation, (c) Experimental stress-strain curve (Figure in (c) reprinted from \cite{zhang2022kink} with permission from \textit{Elsevier}).}
    \label{fig:stress-strain-curve-comparison}
\end{figure}

Fig.~\ref{fig:stress-strain-curve-comparison} (b,c) shows the true and engineering stress-strain curve obtained from our simulation and the experimentally obtained stress-strain curve from \cite{zhang2022kink}, \add{and it is clear that the numerically obtained stress-strain curves do not match with the experimental ones}. The yield strength obtained from our simulations is between 0.45 - 0.5 GPa, while in the experimental results from \cite{zhang2022kink}, it is close to 1.0 GPa. The strain-hardening response is more prominent in our numerical result as compared to the experimental observation, where after yielding it stays at a more-or-less constant level of stress. However, in the current work or our past works like \cite{arora2020unification, arora2022mechanics}, we have not employed any fitting of material parameters to obtain the numerical results that are in qualitative or quantitative agreement with experimental observations. This is discussed in a broader context in Section \ref{sec:comparision_b/w_sgp_mfdm}, where we compare \add{strain gradient plasticity} (SGP) based models and MFDM for modeling problems in mesoscale plasticity.

\begin{figure}[htbp]
    \centering
    \subfloat[][]{
    \includegraphics[scale=0.8]{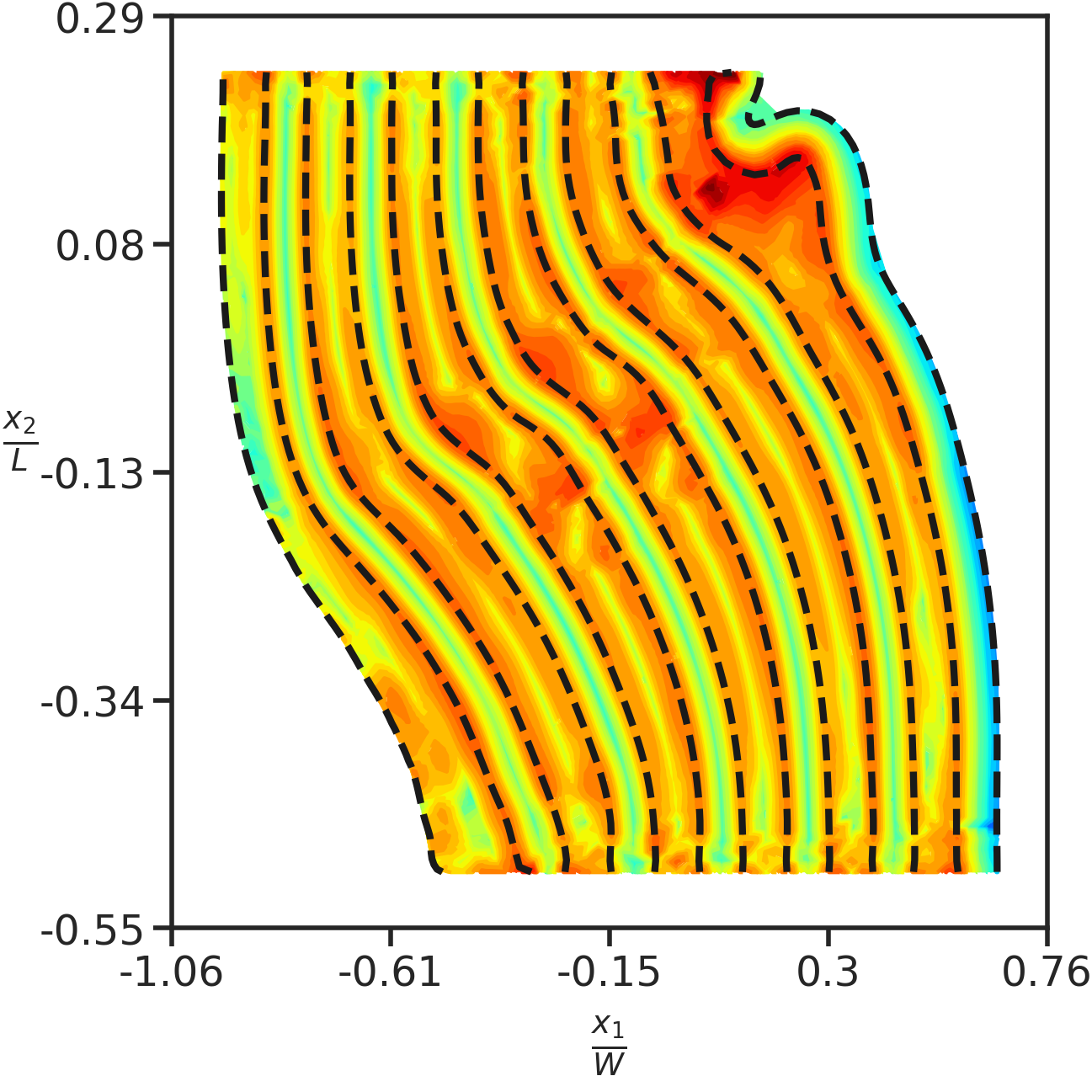}}
    \subfloat[][]{
    \includegraphics[scale=0.8]{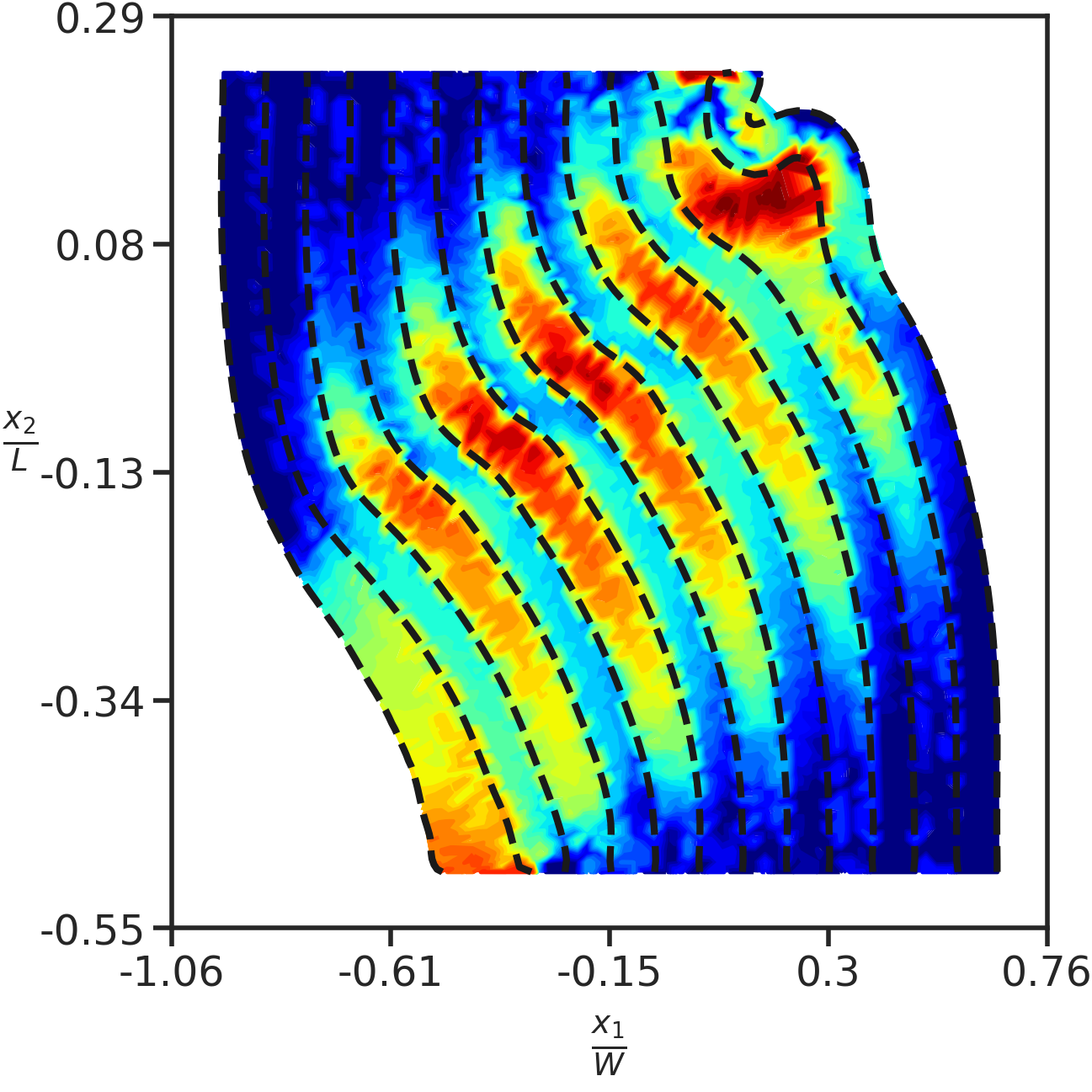}} \quad
    \subfloat{
    \includegraphics[scale = 0.7]{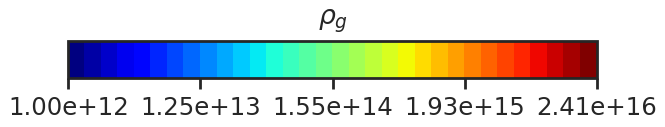}} \quad 
    \subfloat{
    \includegraphics[scale = 0.7]{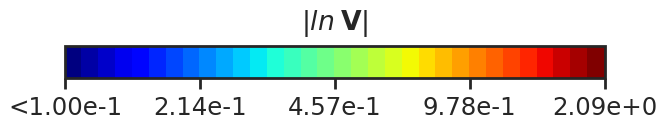}}
    \caption{(a) Field plot of GND density $\rho_g = |\bfalpha|/b \: (m^{-2})$ (b) $|ln \bfV|$ field plot, both at $|e_T| = 23\%$ and $|e_N| = 25.9\%$. The dashed black lines in both Figures show the deformed lines of the interface between Cu and Nb layers in the current configuration.}
    \label{fig:kink_band_formation}
\end{figure}

Fig.~\ref{fig:kink_band_formation} shows the field plot of geometrically necessary dislocation (GND) density ($\rho_g = |\bfalpha|/b \: (m^{-2})$) and the norm of logarithmic strain tensor $|ln \bfV|$. Here, $\bfV$ is obtained from the polar decomposition of the deformation gradient tensor ($\bfF = \bfV \bfR$), and $\bfF$ is defined with respect to the initial configuration at $t=0$. GNDs get generated in the domain due to the inhomogeneity in the material response after yielding from the material inhomogeneity of the laminate structure in NMLs, and the imposition of jump condition on plastic strain rate at the interfaces. 

\begin{figure}[htbp]
    \centering
    \includegraphics[scale = 0.38]{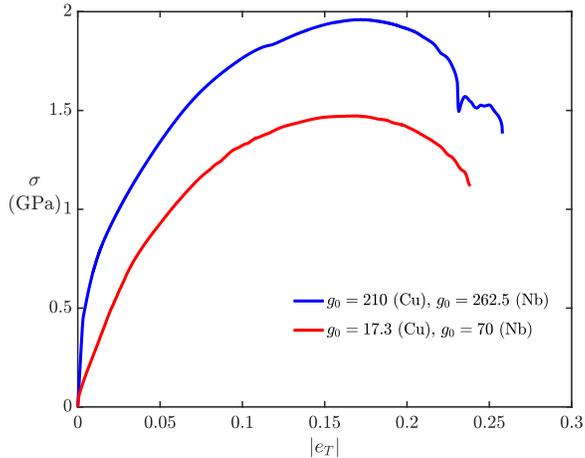}
    \caption{Comparison of true stress-strain curve for different $g_0$, and with $g_s = 9 g_0$ in both cases.}
    \label{fig:g_0_different}
\end{figure}

It is evident from Fig.~\ref{fig:kink_band_formation} that kink-band forms in accordance with the experimental observations in \cite{nizolek2015enhanced, zhang2022kink}. It is evident that the deformation field localizes within the band, and the kink band is inclined (not perpendicular) to the layering in NML as shown in Fig.~\ref{fig:schematic_experimental_kink_band}. Also, as per Fig.~\ref{fig:kink_band_formation} (b) that the principal stretches of $\bfV$ are maximum within the kink band formed, due to intense deformation in the band region. 

The inclination of the kink band shown in Fig.~\ref{fig:kink_band_formation} with respect to the layer normal (in the initial configuration) i.e. $\beta$ (shown in Fig.~\ref{fig:schematic_experimental_kink_band} (a)) is estimated roughly to be $32^{\circ}$, and curiously it is close to $\beta = 35^{\circ}$ reported in the experimental work in \cite{zhang2022kink}. \add{The angle $\beta$ of the inclination of the kink band is calculated based on the actual deformed coordinates of the kink band boundary (note that the aspect ratio of the plot box shown in \mbox{Fig.~\ref{fig:kink_band_formation}} is close to $1:1$).} 

Fig.~\ref{fig:g_0_different} shows the comparison of the stress-strain response with different values of $g_0$ (initial yield), but with the same ratio between the saturation yield strength and initial yield i.e.~$g_s = 9 g_0$. As expected, the initial yield strength for the NML is less, with reduced values of $g_0$, as compared to the experimentally observed yield limit \cite{zhang2022kink}, but the ultimate strength, in this case, is closer to the experimental value. Importantly, we observe the formation of kink bands with similar shapes in our numerical simulations for smaller values of $g_0$ and $g_s$, as well. 

\begin{figure}[htbp]
    \centering
    \subfloat[][]{
    \includegraphics[scale=0.8]{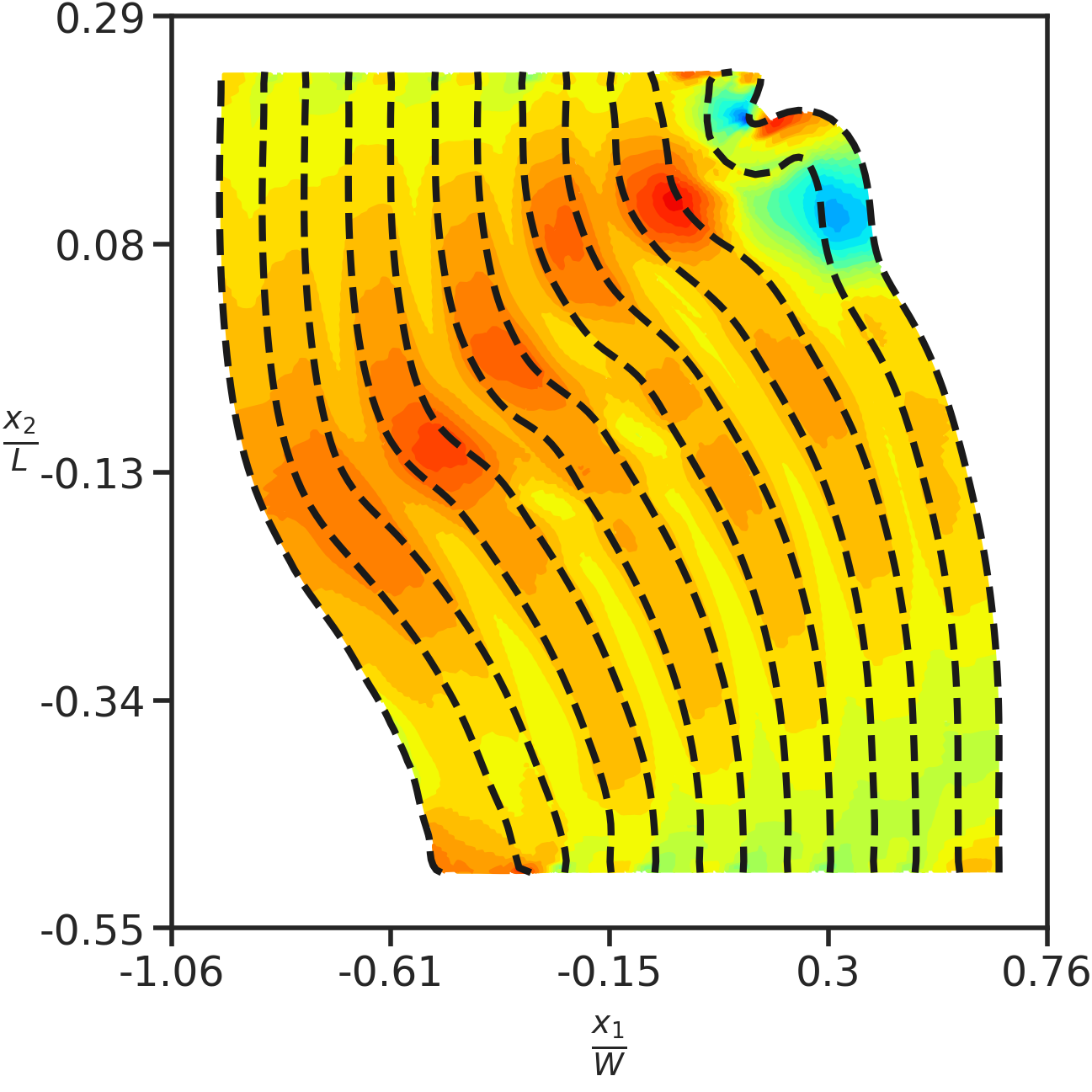}}
    \subfloat[][]{
    \includegraphics[scale=0.8]{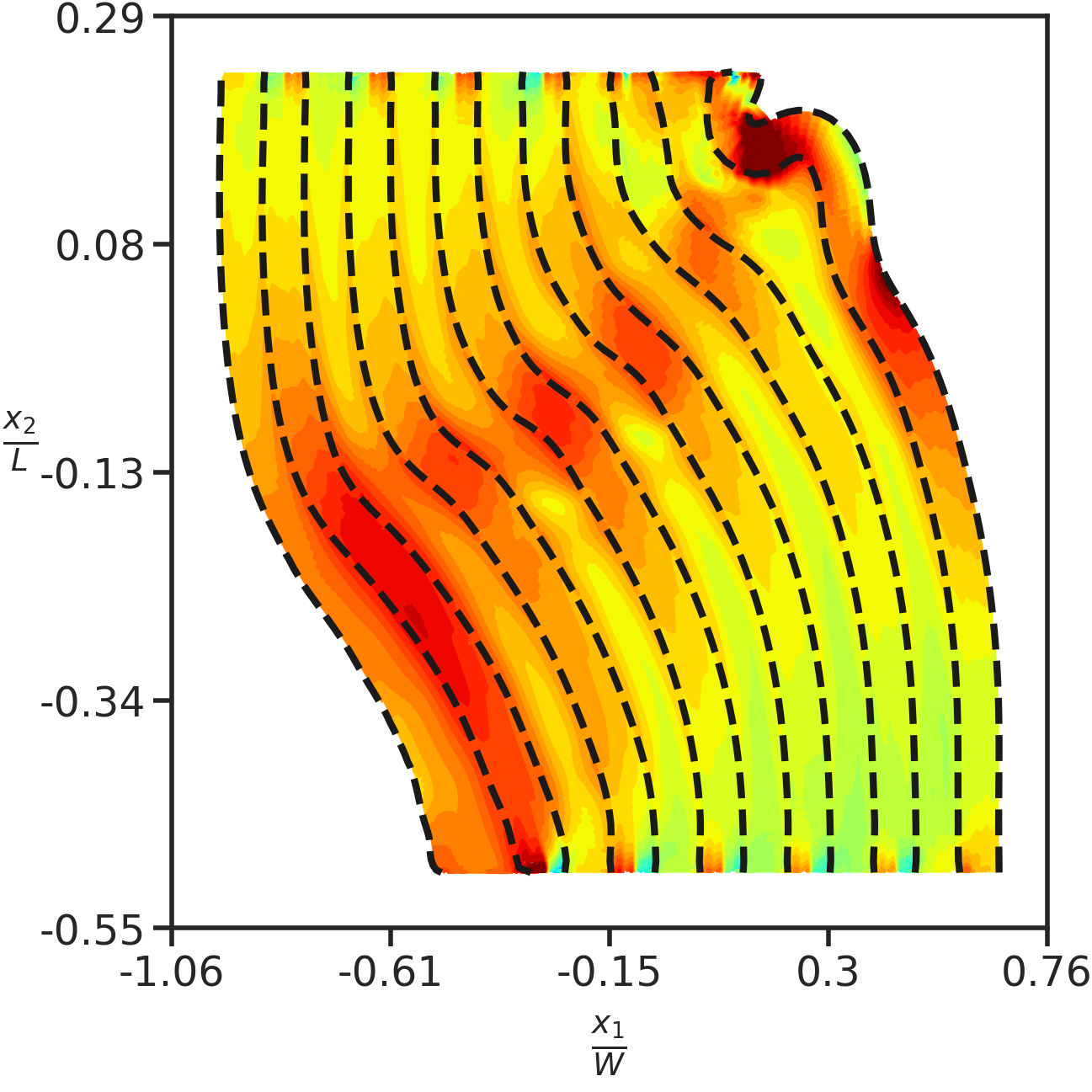}} \quad
    \subfloat[][]{
    \includegraphics[scale = 0.8]{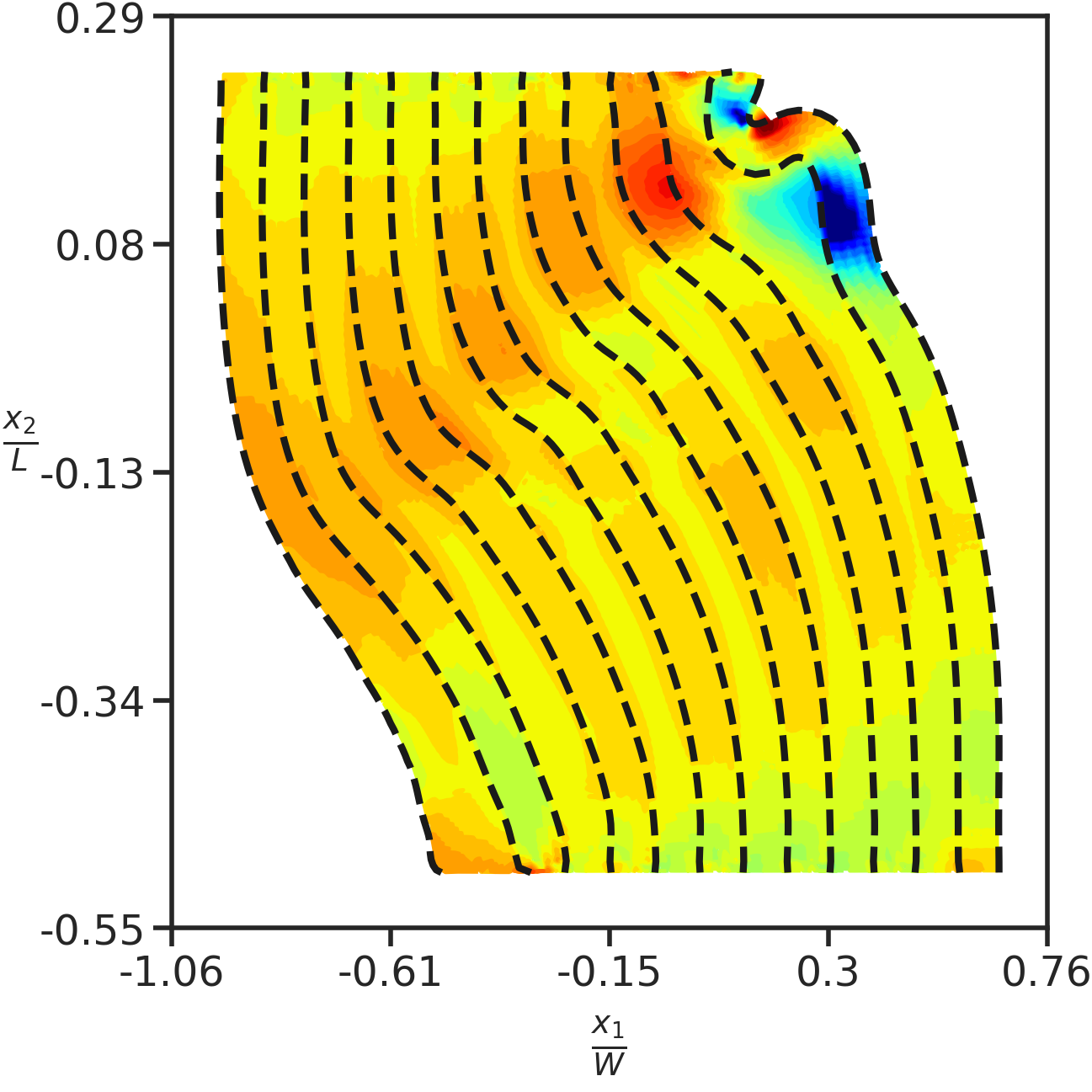}} \\
    \subfloat{
    \includegraphics[scale = 0.8]{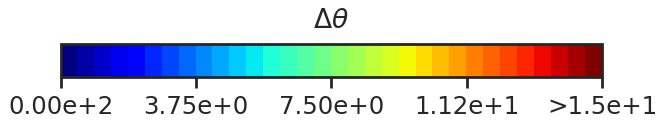}}
    \caption{Rotation field of slip vectors ($\Delta \theta$ in degrees) in the current configuration w.r.t.\ their initial orientation  at $|e_T| = 23\%$ for (a) layer parallel slip vector, (b) $\theta_1$ oriented slip vector, (c) $\theta_2$ oriented slip vector. The dashed black lines in all Figures show the deformed lines of the interface between Cu and Nb layers in the current configuration.}
    \label{fig:rotated_slip_vectors}
\end{figure}

The orientation of the kink band depends \change{upon}{on} the Schmid factors of the 3 slip systems considered, which is the largest for the 2 oblique slip systems. Based on the deformed shape of the NML as shown in Fig.~\ref{fig:rotated_slip_vectors}, the slip direction that is inclined $(\theta_1)$ w.r.t.\ to the layer parallel slip direction becomes the most active slip system based on the Schmid factors in the deformed lattice. However, it is not so clear what breaks the initial symmetry between the double-symmetric slip systems to make the average resolved shear stress on the $\theta_1$ oriented slip plane larger than the $\theta_2$ oriented plane. The absolute value of the averaged Schmid factor (over the whole domain) for the 3 slip systems in the initial configuration is 0 (layer-parallel slip vector) and 0.4665 ($\theta_1$ and $\theta_2$ oriented slip vectors), while the absolute values of the averaged Schmid factors over the domain at $|e_T| = 3.21 \%$ are $0.0021$ (layer parallel slip vector), $0.4739$ ($\theta_1$ oriented slip vector), and $0.4731$ ($\theta_2$ oriented slip vector). At $|e_T| = \: 23\%$, the absolute values of the averaged Schmid factors for the 3 slip systems are $0.1407$ (layer-parallel slip vector), $0.42246$ ($\theta_1$ oriented slip vector), and $0.3357$ ($\theta_2$ oriented slip vector).

The rotation of the deformed slip vectors w.r.t.\ their respective initial orientations is shown in Fig.~\ref{fig:rotated_slip_vectors} at $|e_T| = 23\%$, and it can be deduced that near the kink band region, the rotation of slip vectors is relatively large. The average rotation for the 3 slip vectors is $9.99^{\circ}$ (layer parallel slip vector), $10.27^{\circ}$ ($\theta_1$ oriented slip vector), and $9.64^{\circ}$ ($\theta_2$ oriented slip vector). The negative curl of the inverse elastic distortion tensor is the GND density, and the rotation of lattice vectors is a direct function of the (inverse) elastic distortion tensor. As can be observed from Fig.~\ref{fig:rotated_slip_vectors}, there are large gradients in the rotation field of slip vectors near the kinked region, which results in relatively large magnitudes of GND density in the kink band (refer to Fig.~\ref{fig:kink_band_formation} (a)).

\subsection{Effect of different slip systems, ordering of metallic layers, and slip systems orientations}
\begin{figure}[htbp]
    \centering
    \subfloat[][]{
    \includegraphics[scale=0.8]{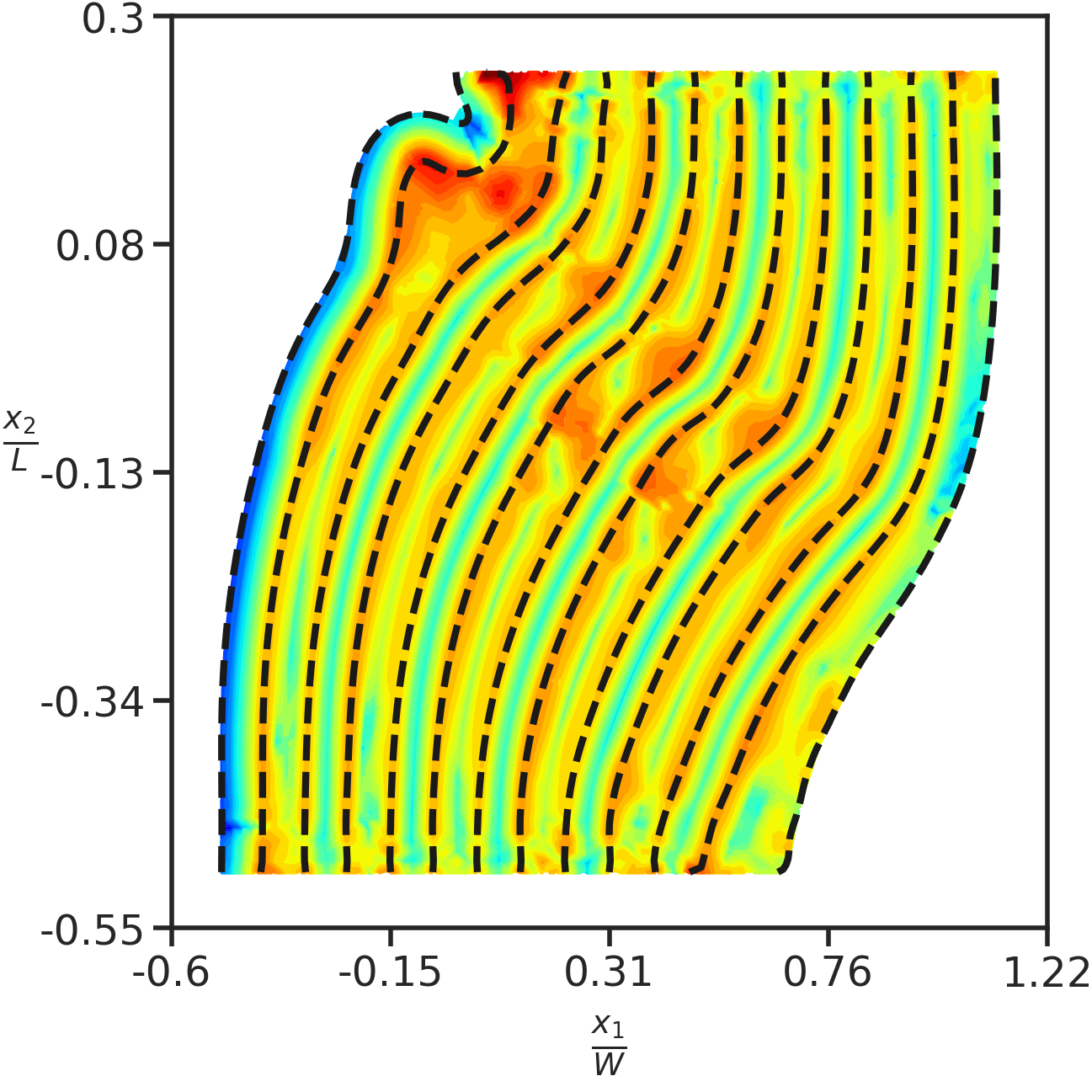}}
    \subfloat[][]{
    \includegraphics[scale=0.8]{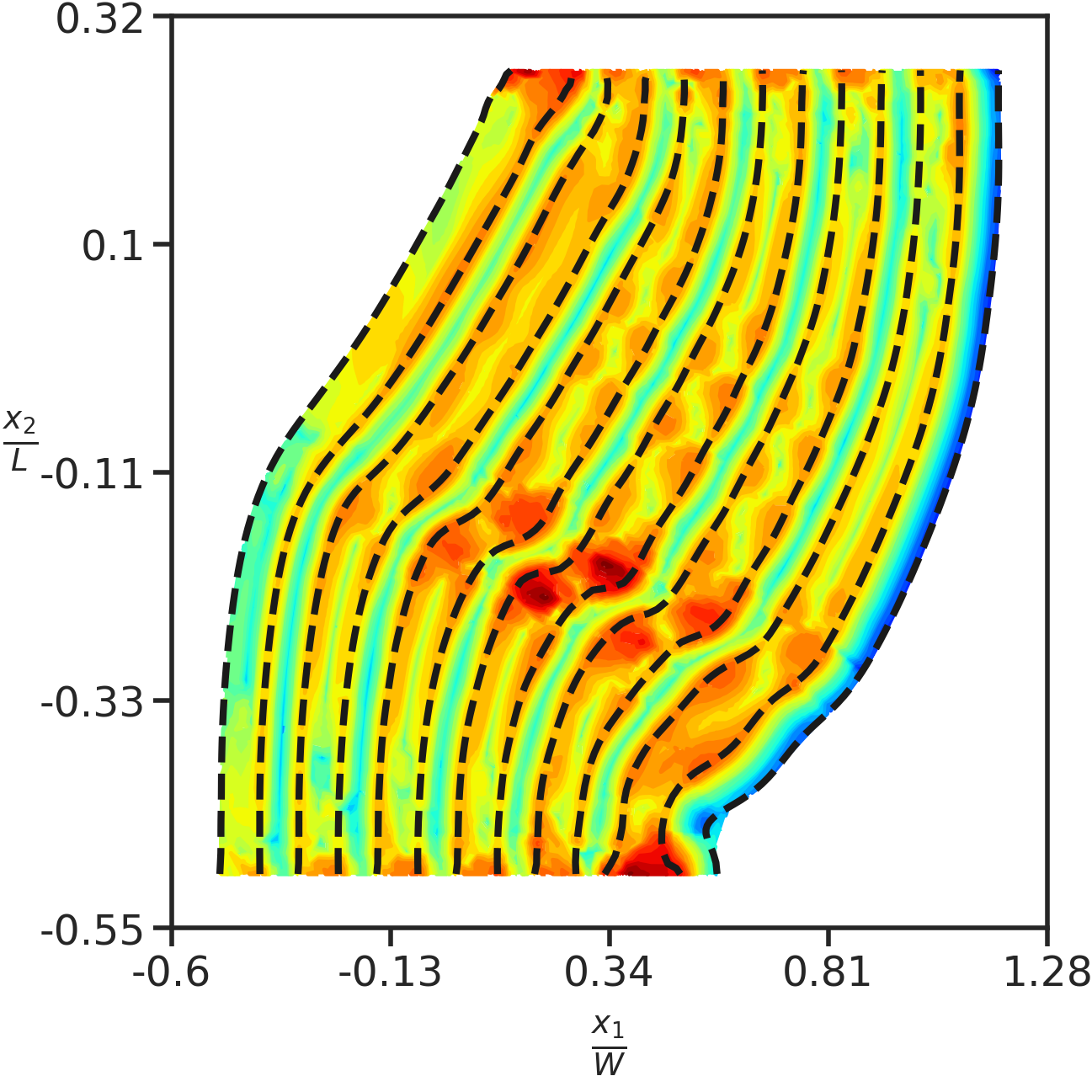}} \quad
    \subfloat{
    \includegraphics[scale = 0.8]{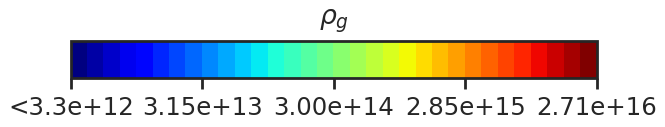}}
    \caption{ Field plot of GND density $\rho_g = |\bfalpha|/b \: (m^{-2})$ (a) after flipping the order of Cu and Nb layers at $|e_T| = 23\%$, (b) after removal of $\theta_1$ oriented slip plane, and at $|e_T| = 20.5\%$, but the order of metallic layers is the same as shown in Fig.~\ref{fig:stress-strain-curve-comparison} (a). The dashed black lines in both Figures show the deformed lines of the interface between Cu and Nb layers in the current configuration.}
    \label{fig:kink_band_formation_flip}
\end{figure}

\begin{figure}[htbp]
    \centering
    \subfloat[][$\rho_g$ field plot (2-slip)]{
    \includegraphics[scale=0.8]{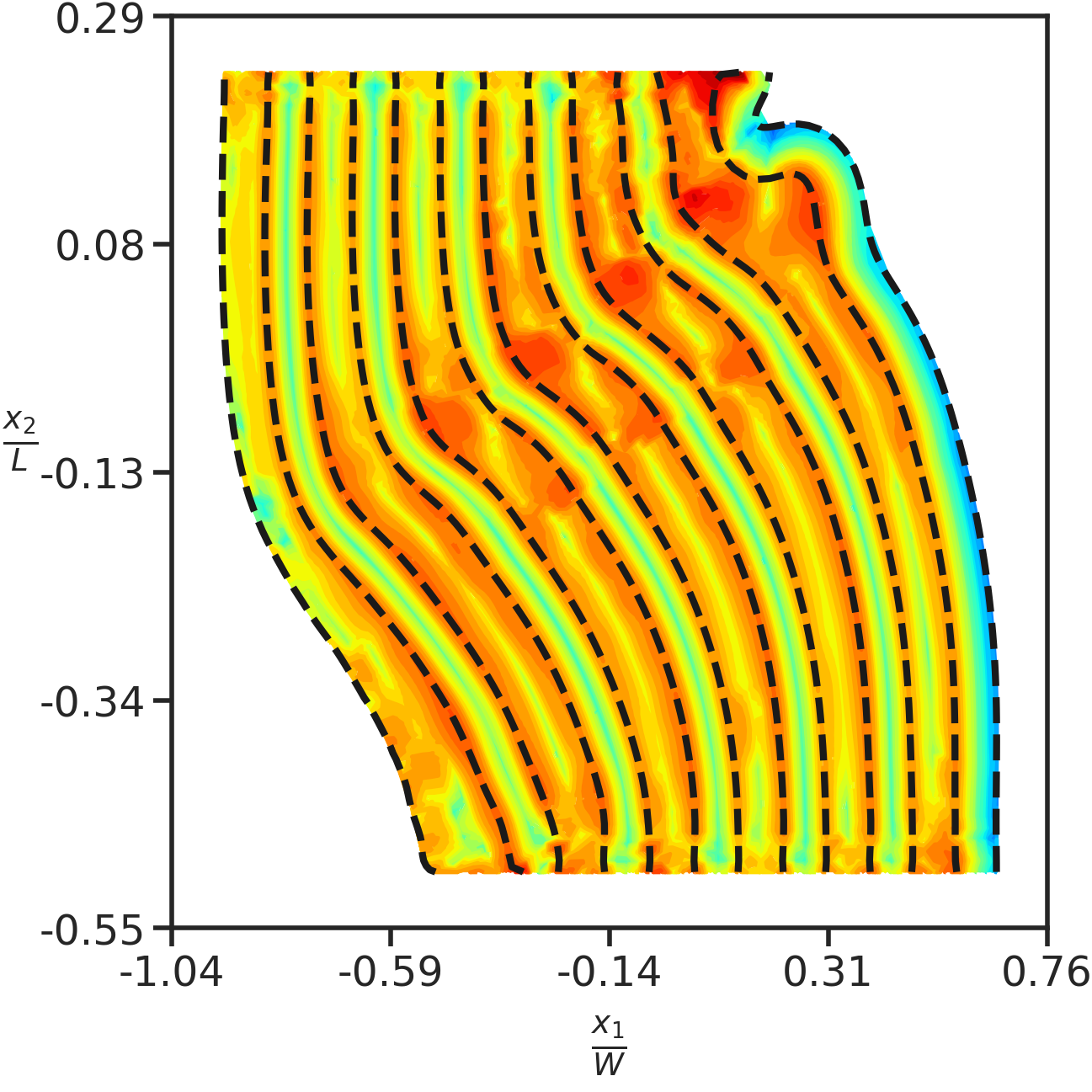}}
    \subfloat[][$\rho_g$ field plot (3-slip)]{
    \includegraphics[scale=0.8]{plots/alpha_jan7_18000.png}} \quad
    \subfloat{
    \includegraphics[scale = 0.8]{plots/colorbar_alpha_jan7_18000.png}}
    \caption{(a) $\rho_g$ field plot for 2-slip system only (b) $\rho_g$ field plot for all 3-slip systems, both at $|e_T| = 23\%$. The dashed black lines in both Figures show the deformed lines of the interface between Cu and Nb layers in the current configuration.}
    \label{fig:layer_parallel_rho_g}
\end{figure}

We report various numerical studies with respect to changing the ordering of metallic layers, the effect of removal of slip vectors, and the slip vector orientations on the kink band formation. 
\begin{itemize}
    \item Flipping the order of Cu and Nb layers in the nano-laminate structure resulted in a change in the orientation of the band as shown in Fig.~\ref{fig:kink_band_formation_flip} (a), due to a geometrical asymmetry in the initial configuration.
    \item Removal of $\theta_1$ inclined slip system, and just considering layer parallel slip system and $\theta_2$ inclined slip system, resulted in the orientation of the band as shown in Fig.~\ref{fig:kink_band_formation_flip} (b). 
    \item Removal of the $\theta_2$ oriented slip plane did not change the orientation of the kink band formed, and it is similar to as shown in Fig.~\ref{fig:kink_band_formation}.
    \item Removal of the layer parallel slip system and just considering the double symmetric slip system with the same orientations, as before, did not change the orientation of the kink band formed, as shown in Fig.~\ref{fig:layer_parallel_rho_g}. 
    \item We consider different orientations of slip vectors ($\theta_1 = \theta_2$)  from  $20^{\circ}$ to $45^{\circ}$ for Cu, and keep the slip vector orientation for Nb to be the same i.e.~$\theta_1 = \theta_2 = \: 45^{\circ}$. Kink bands form for slip vector orientations from $20^{\circ}$ to $30^{\circ}$, but for $35^{\circ}$ to $45^{\circ}$ kink bands do not form, and NMLs undergo more uniform compression. 
\end{itemize}

\begin{figure}[htbp]
    \centering
    \includegraphics[scale=0.38]{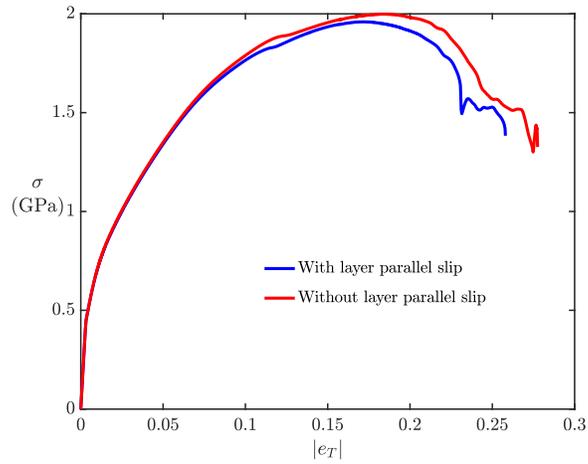}
    \caption{True stress-strain curve with/without accounting for layer-parallel slip system.}
    \label{fig:layer_parallel-ss}
\end{figure}

The stress-strain response with just 2 slip systems is relatively stronger as compared to the 3-slip system case, as shown in Fig.~\ref{fig:layer_parallel-ss}. As concerns, the $\sim$ flat experimental stress-strain response beyond yielding, the availability of more slip systems in realistic modeling of the NML would lead to more plastic straining, but it would be accompanied by more gain in strength as well. So, in general, it is difficult to say with certainty that the availability of more slip systems will result in less strain-hardening, but that is what we observe in our limited simulations reported here.

We further compare the field plots of GND density ($\rho_g$) with/without the layer parallel slip system as shown in Fig.~\ref{fig:layer_parallel_rho_g} at $|e_T| =  23\%$ strain. It is evident that the shape of the kink band and the GND density are similar in the two cases. This is a departure from the results reported in the experimental work by Zecevic et al.~\cite{zecevic2023non}, where it was reported that the layer-parallel slip system is responsible for the evolution of the shape of the kink band. We leave a more realistic investigation of the effects of layer parallel slip systems for future work, where we will be modeling the 3-D geometry of NML and considering all the slip systems in Cu (FCC) and Nb (BCC). 

\add{Another interesting observation is that} the domain-averaged rotation, with respect to the initial orientation, decreases with an increase in the initial orientation angle of the slip vectors from the loading axis, in the range of initial orientations where kink bands form. This shows that in order to accommodate the same level of plastic deformation, slip vectors with smaller angular deviation from the loading axis rotate more on average during plastic deformation.

\subsection{Effect of layer thickness length scale on the formation of kink band}
\begin{figure}[htbp]
    \centering
    \subfloat[][]{
    \includegraphics[scale=0.35]{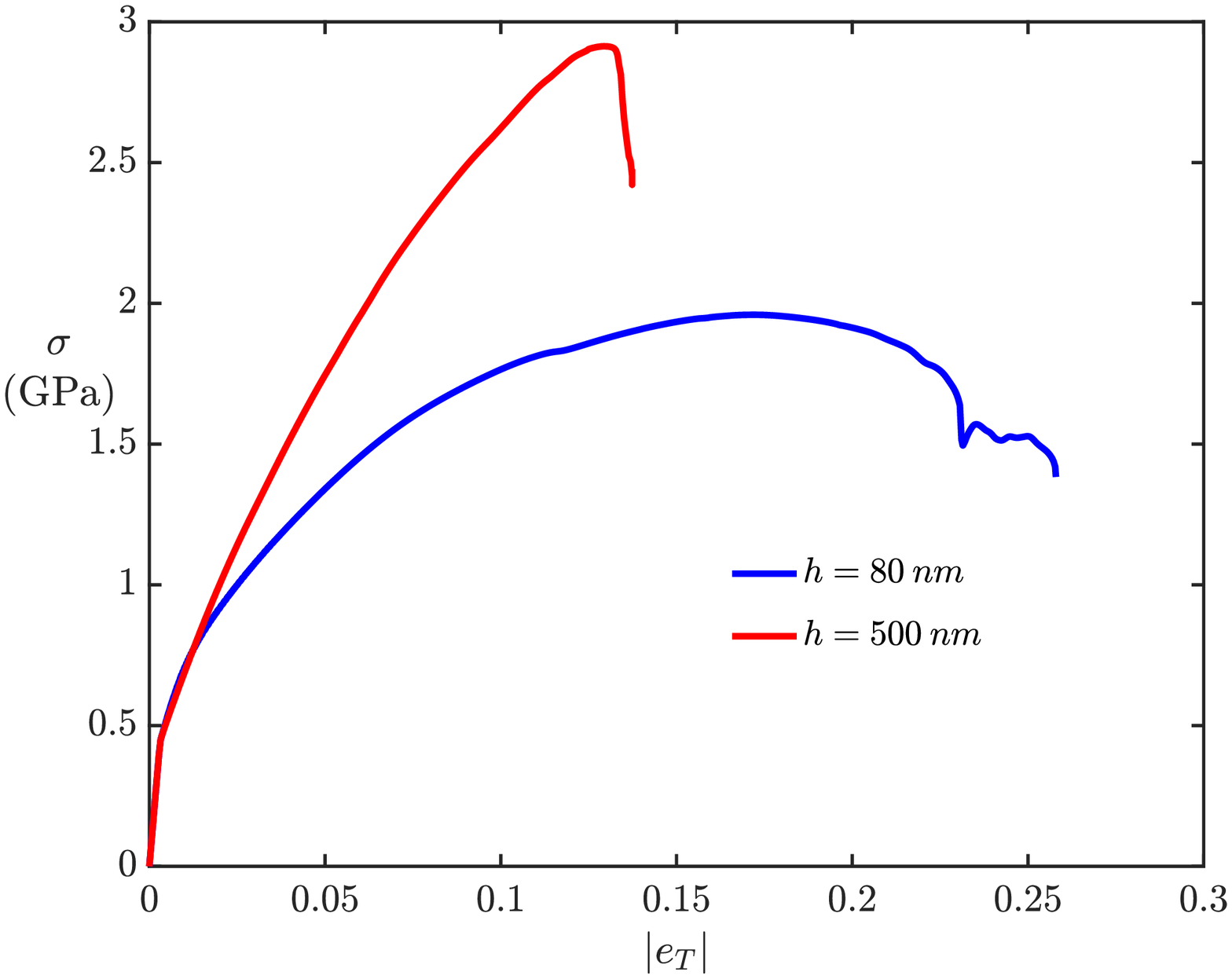}}
    \subfloat[][]{
    \includegraphics[scale=0.8]{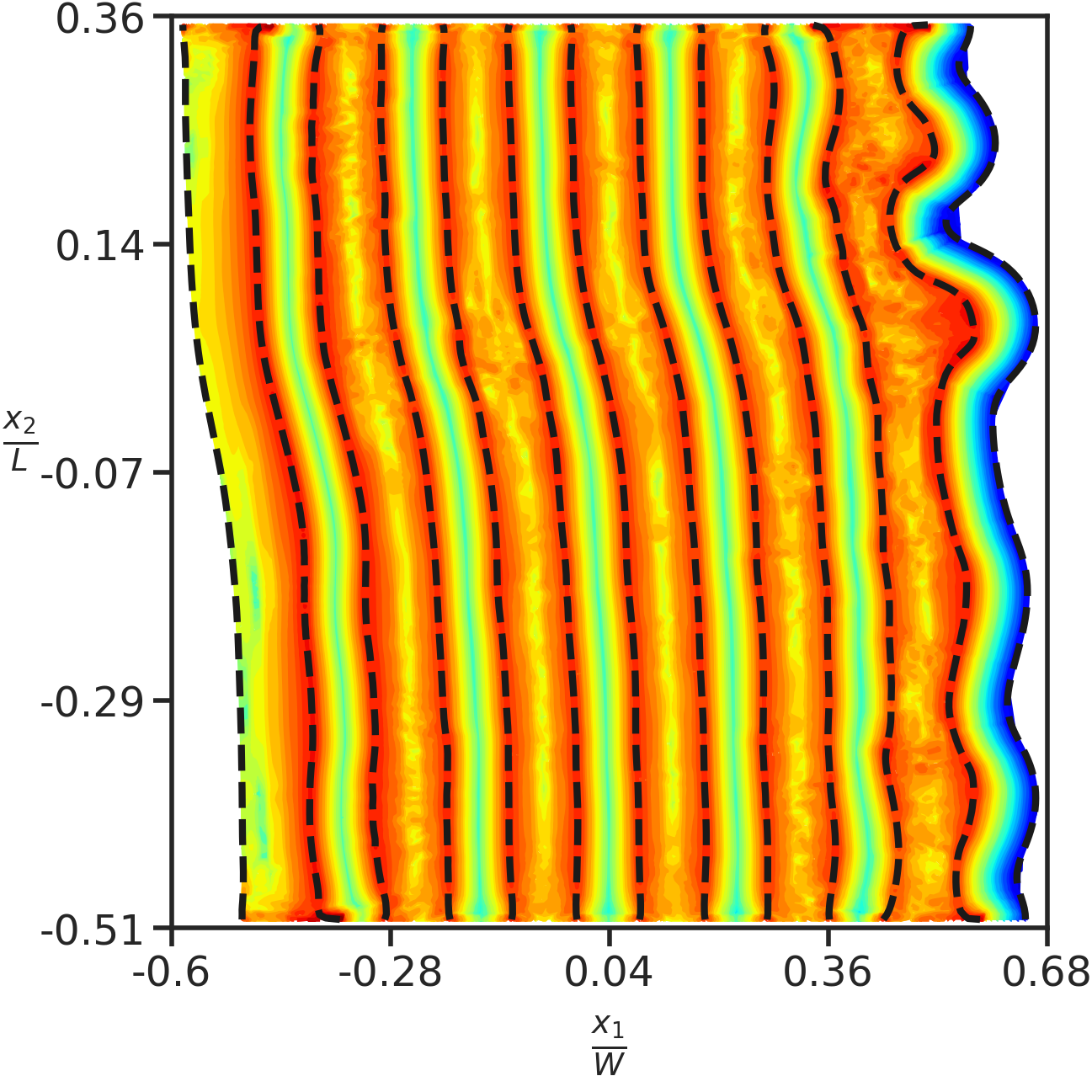}}\quad
    \subfloat{
    \includegraphics[scale=0.8]{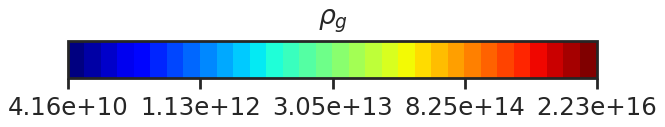}}
    \caption{(a) True stress-strain curve comparison for layer thickness $h = 80 \: nm$ and $h = 500 \: nm$, (b) $\rho_g$ field plot at $|e_T| = 13.63\%$ for layer thickness $h=500 \: nm$. The dashed black lines in Figure (b) show the deformed lines of the interface between Cu and Nb layers in the current configuration.}
    \label{fig:layer_thickness}
\end{figure}

We investigate the effect of layer thickness on the formation of kink bands in NMLs. It was experimentally observed in the work by Nizolek et al.~\cite{nizolek2015enhanced} that kink bands do not form and homogeneous deformation is observed for layer thickness greater than $250 \: nm$.
We run our simulation with the same parameters as stated in Table \ref{table:material_properties}, and take the layer thickness $l = 500 nm$, and the NML is composed of a total of 12 layers. Hence, the width of the NML composite is $W = 6 \: \mu m$, and keeping the same aspect ratio of 1:2, the length of the NML composite is $L = 12 \: \mu m$. It can be deduced from the deformed shape of NML (shown in Fig.~\ref{fig:layer_thickness} (b)) that kink bands do not form with larger layer thickness, and it is getting compressed homogeneously without the formation of any kink bands. 

\subsection{Conventional plasticity within our model}
\begin{figure}[htbp]
    \centering
    \subfloat[][Stress-strain curve]{
    \includegraphics[scale=0.35]{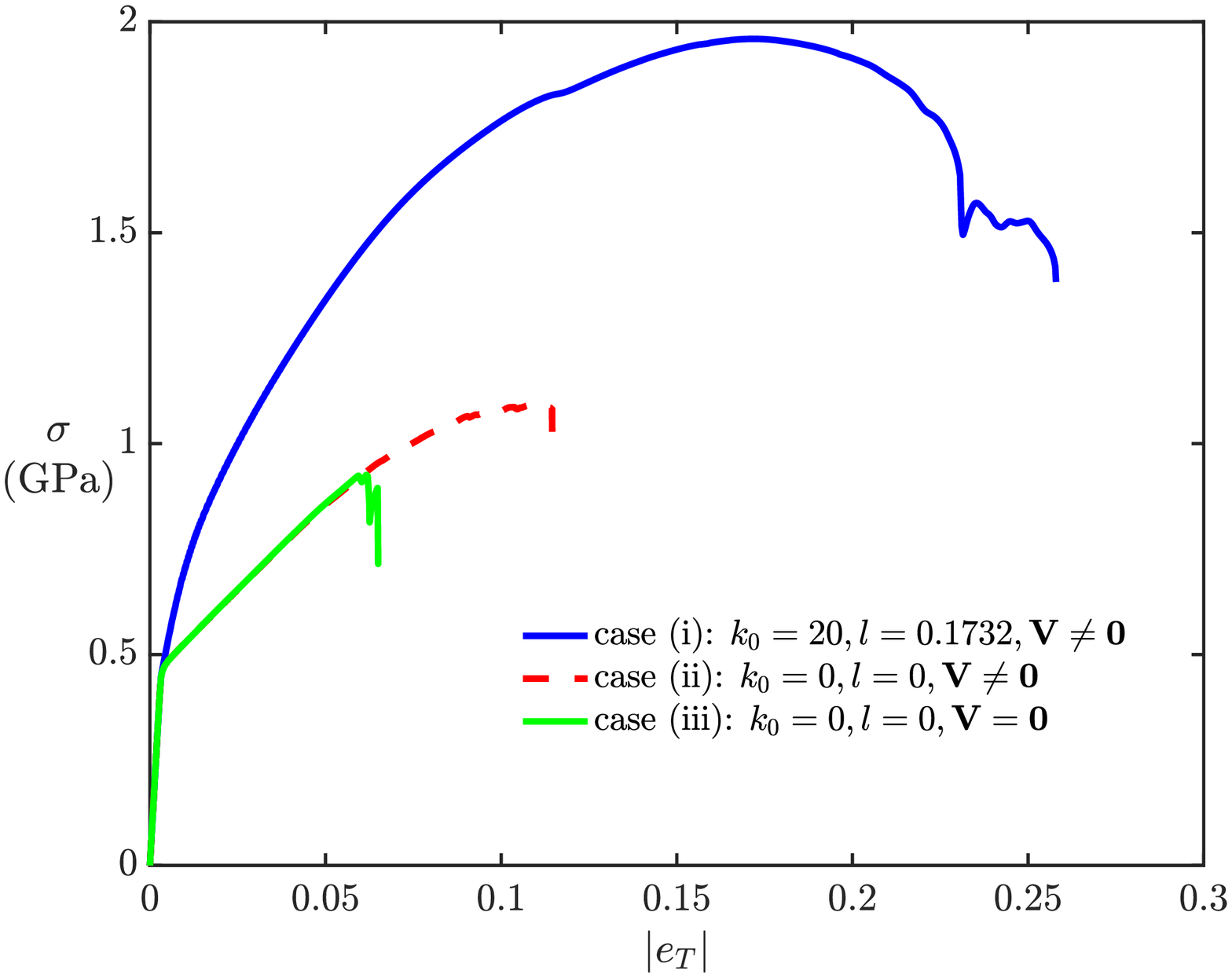}}
    \subfloat[][$\rho_g$ field plot]{
    \includegraphics[scale=0.85]{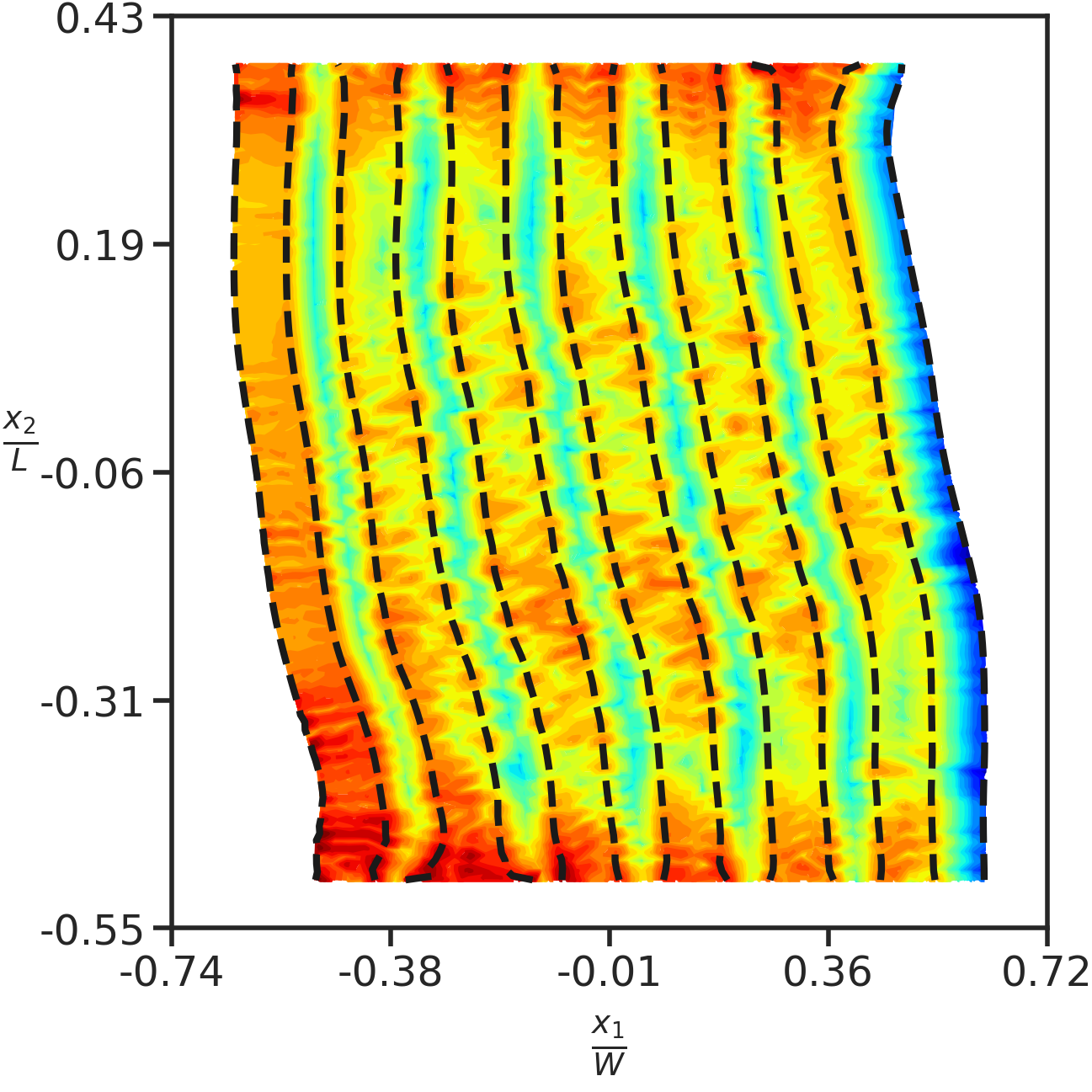}} \quad
    \subfloat{
    \includegraphics[scale = 0.8]{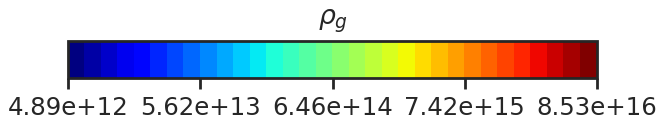}}
    \caption{(a) True stress-strain curve comparison for the 3 cases considered, (b) $\rho_g$ field plot at $|e_T| = 11.3\%$ for case (ii): $k_0 = 0, \: l = 0, \: \bfV \neq \bf0$. The dashed black lines in Figure (b) show the deformed lines of the interface between Cu and Nb layers in the current configuration.}
    \label{fig:conv_plasticity}
\end{figure}

In this section, we investigate the response of NML in three cases. For case (i), we take the usual parameters ($k_0 = 20, l = 0.01732, \bfV \neq \bf0$), while for case (ii) the parameters are  ($k_0 = 0, l = 0, \bfV \neq \bf0$), and for case (iii) the parameters taken are ($k_0 = 0, l = 0, \bfV = \bf0$). Physically, case (ii) corresponds to just Voce's law-based hardening (without the effects of GND hardening) and no plastic strain rate due to mesoscopic core energy. For case (iii), in addition to case (ii), there is no plastic straining due to the motion of GNDs, and this case is closest to conventional plasticity within our unified crystal plasticity and dislocation mechanics framework. As expected, with no hardening from GNDs accounted for in case (ii) and case (iii), we obtain a weaker strain-hardening response, as compared to case (i). It was observed in numerical simulations that kink bands do not form for case (ii) and case (iii), as shown in Fig.~\ref{fig:conv_plasticity} (b),  which shows the GNDs density field plot for case (ii) at $|e_T| = 12\%$ true strain.

\subsection{NML subjected to layer-perpendicular compression}

\begin{figure}[htbp]
    \centering
    \includegraphics[scale = 0.4]{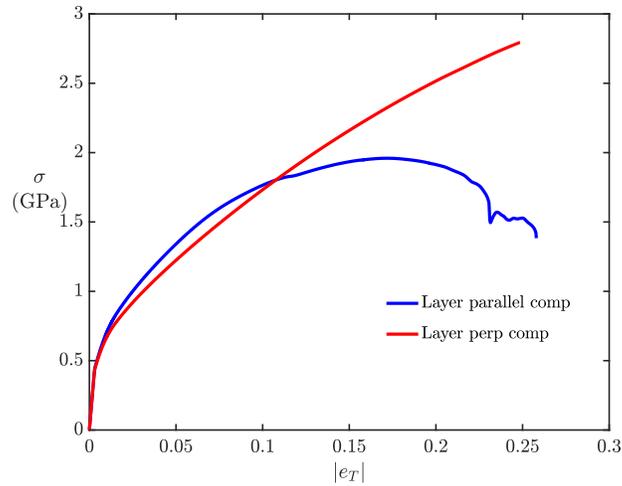}
    \caption{True stress-strain curve when the layer direction is aligned perpendicular or parallel to the compression loading direction.}
    \label{fig:layer_perp_compression}
\end{figure}

\begin{figure}[htbp]
    \centering
    \includegraphics[scale = 0.7]{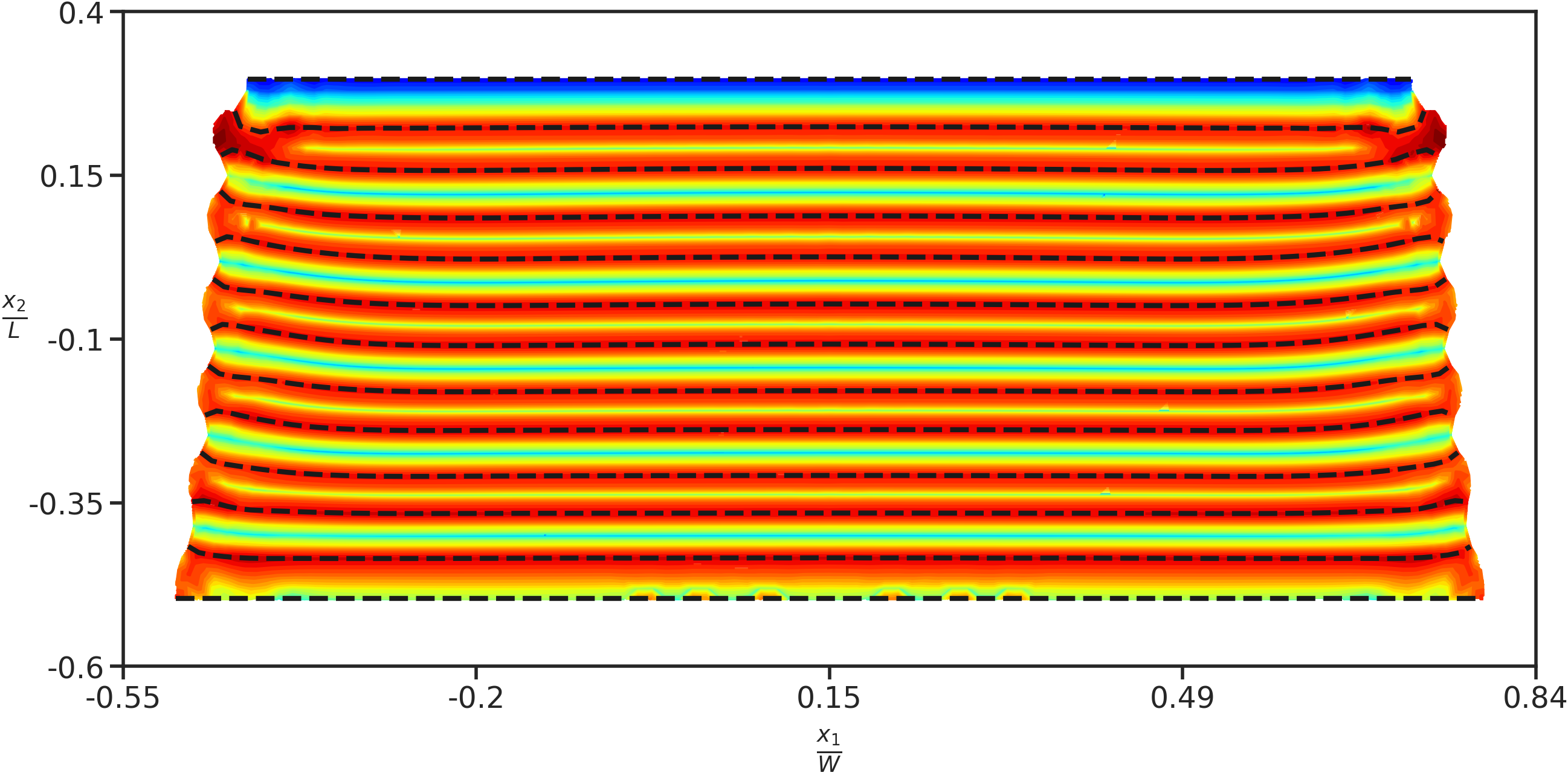} \\
    \includegraphics[scale=0.8]{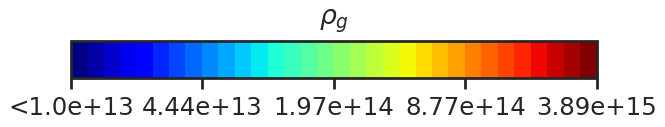}
    \caption{$\rho_g$ field plot at $|e_T| = 18.23\%$ strain, when the layer direction is aligned perpendicular to the compression loading direction. The dashed black lines in the Figure show the deformed lines of the interface between Cu and Nb layers in the current configuration.}
    \label{fig:alpha_layer_perp_compression}
\end{figure}

In this section, the layer direction of NML is aligned perpendicular to the compression loading direction. As observed in the experiments \cite{beyerlein2013interface}, kink bands do not form for layer perpendicular compression and the deformation of the NML is shear-driven parallel to the layer direction. This results in material flow under shearing within the individual layers as shown in Fig.~\ref{fig:alpha_layer_perp_compression}. For the layer perpendicular micropillar compression of NMLs in \cite{beyerlein2013interface}, it was reported that diagonally inclined shear bands are formed, from the left or right boundary to the right/left boundary. We leave such studies for future work where we model the slender micropillar shape of the NML using a plane strain idealization with the undulating interfaces. Up to the rotation of the layers towards the loading axis, similar behavior is observed in the work \cite{arora2022mechanics}.

\subsection{Comparison of \change{SGP}{strain gradient plasticity (SGP)} models and MFDM} \label{sec:comparision_b/w_sgp_mfdm}
\begin{itemize}
    \item The numerical results that we have obtained are with continuous plastic flow across the interfaces, imposing continuity of certain components of plastic strain rate across the interface. Moreover, we see the accumulation of GND density across the interface (refer to Fig.~\ref{fig:kink_band_formation} (a)), without having to do constrained plastic flow across the interface, which models the impenetrable interfaces in MFDM framework, as discussed in \cite{puri2011mechanical}. This is unlike the recent work by Zecevic et al.~\cite{zecevic2023non}, where `micro-hard' boundary conditions are implemented on the interfaces for their numerical study. From the current state-of-the-art of SGP models \cite{kuroda2021constraint, zecevic2023non} there appears to be a high degree of indeterminacy in the nature of boundary/interface conditions to be imposed, with a significant impact on model predictions.
    
    Using `micro-hard' boundary conditions within SGP models would make it difficult to reproduce drastically different scalings in the micropillar compression experiments for two different configurations ($90^{\circ}$ and $45^{\circ}$ oriented metal thin film with respect to compression loading axis), as experimentally observed in \cite{mu2016dependence}. In the work of Kuroda et al.~\cite{kuroda2021constraint}, the plastic constraints on the metal-ceramic interface are relaxed beyond a certain level of plastic strain gradient on the interface/boundary, in order to obtain the same scalings as experimentally observed. However, our model is able to reproduce similar scalings as compared to experimental ones, without any ad-hoc modifications to the boundary conditions, as shown in \cite{arora2022mechanics}. Moreover, our model also predicts the formation of kink bands in the compression of NMLs, as shown in this current (first) simplified study, again with no special fitting of material parameters or changes to the structure of the theory (which, of course, includes the nature of boundary and interface conditions). The jump conditions of our model have also been shown to be successful in studies of texture evolution and recrystallization, among others \cite{mach2010continuity, fressengeas2020continuum, su2021predicting}.
    \item Our model is not able to reproduce size dependence on the initial yield strength, as observed in the micropillar compression experiments of single crystal pure Ni \cite{uchic2004sample}; to our knowledge, SGP-based theories have not modeled these experiments either. Most SGP theories are able to predict a significant size effect at initial yield in the presence of boundary constraints or in the presence of inhomogeneous deformation, without being able to fundamentally distinguish boundary constraints arising from the kinematics of dislocation slip \cite{arora2020unification,arora2022mechanics}.
    
    \item SGP studies of micropillar confined thin film plasticity and kink banding \cite{kuroda2019simple, kuroda2021constraint, zecevic2023non}, usually employ a $\sim 0$ value of work hardening to reproduce experimental behavior, raising the question of accurate representation of macroscopic behavior by the models. Moreover, the SGP studies of \cite{zecevic2023non, nicola2005effect} employ several variants of the core energy  function and show a sensitive and significant dependence on stress-strain response to such a choice. In our work, no such choices are required, and the work hardening rates we use in our strength evolution ensure, \add{at least in the confined thin film problem and for modeling macroscopic response, that a physically appropriate mechanical response is obtained. For the present kink-band problem, the obtained stress-strain response (without any fitting) is unsatisfactory compared to the experiment (but no worse than the SGP result of \mbox{\cite{zecevic2023non}}), and this is an issue that requires further work.}
    
\end{itemize}

\section{Conclusion} \label{conclusion}
This work reports a preliminary study of the modeling of kink-banding in NML within the MFDM framework. We report reasonable success, which further strengthens the ability of the MFDM framework to model important and  challenging problems in the field of modern plasticity. Our simulations of a minimalistic model recover important experimental results in the compression of NMLs such as:
\begin{itemize}
    \item dependence of kink-band formation on the layer thickness.
    \item kink bands do not form when the layer direction is aligned perpendicular to the compression loading direction, and there is shear-driven deformation within the layers of NML. 
\end{itemize}
At this time we are unable to provide a detailed analysis of the location of the kink bands in our numerical simulations. Rough speculation in this regard is as follows: the bands cannot cross the top and bottom boundaries because of the applied velocity boundary conditions. The corners of the sample result in stress concentrations due to geometric and changes in velocity boundary conditions on intersecting external boundaries. These stress concentrations may be the source of the location of the bands with the orientation governed by the most active slip systems. Of course, there are subtle details of the effects of the GND evolution that definitely affect the localization of the band and the layer length scale effects. Such a detailed understanding of the localization of the kink band is a fundamental challenge for future work as is the modeling of the full 3-D geometry of NML samples and the crystallography of actual slip systems present in Cu and Nb.

The prediction of experimentally observed kink bands in NMLs \cite{nizolek2015enhanced, zhang2022kink} along with the micropillar confined thin film plasticity \cite{mu2014thickness, mu2016dependence} form a discerning, critical set of key experimental tests of mesoscale plasticity theories, to be accomplished without ad-hoc model modifications on a case-by-case basis. The MFDM model has been shown to be in qualitative accord with these experimental tests, building confidence in its adequacy as a model for the intricate phenomena involved in mesoscale plasticity.

\section*{Acknowledgments}
This work was supported by the grant NSF OIA-DMR $\#2021019$.

\bibliographystyle{alpha}\bibliography{paper_template.bib}
\end{document}